\documentclass[prd,tightenlines,nofootinbib,superscriptaddress]{revtex4}

\usepackage{pre_all}
\usepackage{pre_CS}
\allowdisplaybreaks

\hypersetup{
pdfstartview = {FitH},
}
\hypersetup{
	colorlinks=true,         
	linkcolor=brown,          
	citecolor=red,        
	urlcolor=blue            
}

\begin{document}

\title{Deficit Angles in 4D Spinfoam with Cosmological Constant: (Anti) de Sitter-ness and More}

\author{{\bf Muxin Han}}\email{hanm@fau.edu}
\affiliation{Department of Physics, Florida Atlantic University, 777 Glades Road, Boca Raton, FL 33431, USA}
\affiliation{Institut f\"ur Quantengravitation, Universit\"at Erlangen-N\"urnberg, Staudtstr. 7/B2, 91058 Erlangen, Germany}

\author{{\bf Qiaoyin Pan}}\email{qpan@fau.edu}
\affiliation{Department of Physics, Florida Atlantic University, 777 Glades Road, Boca Raton, FL 33431, USA}

\date{\today}

\begin{abstract}
This paper investigates the critical behaviors of the 4-dimensional spinfoam model with cosmological constant for a general 4-dimensional simplicial complex as the discretization of spacetime. We find that, at the semi-classical regime, the spinfoam amplitude is peaked at the real critical points that correspond to zero deficit angles (modulo $4\pi\mathbb{Z}/\gamma$) hinged by internal triangles of the 4-complex. Since the 4-simplices from the model are of constant curvature, the discrete geometry with zero deficit angle manifests a de Sitter (dS) spacetime or an anti de Sitter (AdS) spacetime depending on the sign of the cosmological constant fixed by the boundary condition. The non-(A)dS spacetimes emerge from the complex critical points by an analytic continuation to complex configurations. 
\end{abstract}

\maketitle

\tableofcontents

\section{Introduction}

In the research area of Loop Quantum Gravity (LQG), the understanding of quantum gravity without a cosmological constant is far deeper than that of quantum gravity with a non-vanishing cosmological constant $\Lambda$. It is because the introduction of $\Lambda\neq 0$ brings complexity to the quantum geometry construction and it is commonly believed (and exemplified in lower dimensional LQG) that new mathematical tools need to be applied. 
In the covariant, also called the spinfoam, LQG approach, this is especially the case. 
In 3+1 dimensions (4D), the investigations of spinfoam models are mostly on those with $\Lambda=0$, among which the most studied one is the EPRL model \cite{Engle:2007wy}. 
It is hoped that an adequate understanding of spinfoam model(s) with $\Lambda=0$ can shed light on constructing a $\Lambda\neq0$ version. 
In this paper, inversely, we present a result from the spinfoam model with $\Lambda\neq0$ that helps resolve an ambiguity in the EPRL model.  

Our analysis is based on the spinfoam model introduced in \cite{Han:2021tzw}. It describes Lorentzian 4D quantum gravity with a either positive or negative cosmological constant, which is taken as a global coupling constant and whose sign depends on the boundary geometrical condition. One of the advantages of this spinfoam model compared to other existing ones with $\Lambda\neq 0$ (\eg \cite{Noui:2002ag,Han:2010pz,Fairbairn:2010cp,Haggard:2014xoa,Haggard:2015nat,Haggard:2015sl}) lies in that it not only illustrates discrete curved geometry in its semi-classical regime but also manifests the expected finiteness in the amplitude. 
In this paper, we give a complete description of the amplitude in this spinfoam model for a general 4-complex and show that it retains the finiteness property. 
We then focus on the semi-classical regime of the amplitude and analyze its asymptotic behaviour and geometrical interpretation. As desired, the peak of the amplitude can be interpreted as 4-simplices glued together by identifying boundary geometries to form a 4-complex. This is consistent with the preliminary result in the original paper \cite{Han:2021tzw}. 

At the semi-classical regime of the amplitude, an action of configuration variables can be constructed. Of particular interest in this paper are the equations of motion concerning the internal spins $j_f$'s that dress the internal triangles of the 4-complex. 
In the EPRL model, the use of Poisson resummation for $j_f$'s in the semi-classical approximation of the amplitude gives infinite sums in the following form (see e.g. \cite{Han:2021kll})
\be
\cZ_{\text{EPRL}}=\sum_{\{j_f\}\in \N/2}\prod_f\cA(j_f)\int\rd \mu(X)\, e^{\sum_f j_f F_f(X)} = \sum_{\{u_f\}\in\Z} \int \prod_f\cA(j_f)\rd (2j_f) \int\rd \mu(X)\, e^{\sum_f j_f \lb F_f(X) + 4\pi i u_f\rb }\,,
\label{eq:EPRL_amplitude}
\ee
where $X$ describes all spinfoam integration variables other than $j_f$'s, $\mu(X)$ the collection of their measures and $F_f(X)$ a function on these variables. By the stationary phase analysis in the large-$j$ regime, the real critical point\footnote{The real critical point is inside the integration domain understood as a real manifold. The complex critical point is in the complexified integration domain and in general away from the real integration domain.} describes the deficit angles $\varepsilon_f$'s hinged by the internal triangles \cite{Barrett:2009gg}. However, in this model, it is not clear how many $u_f$'s contribute to the critical point, so in principle, one needs to perform the stationary phase analysis for every $\{u_f\}$, although there are numerical evidences that only one $u_f$ contributes in simple models \cite{Han:2021kll}. The deficit angles may take values $\varepsilon^{\text{EPRL}}_f=4\pi\Z/\gamma$ \cite{Bonzom:2009hw,Han:2013hna}, where $\gamma$ is the Barbero-Immirzi parameter. The critical point with $\varepsilon^{\text{EPRL}}_f=0$ corresponds to the smooth flat geometry, since the 4-simplices are endowed with the flat geometry. 

We observe that the spinfoam model with $\Lambda\neq0$ involves a similar Poisson resummation, but there is only {\it one} $\{u_f\}$ corresponding to the dominant contribution in the semi-classical approximation. In contrast to \eqref{eq:EPRL_amplitude}, the semiclassical amplitude receives the dominant contribution from only one term:
\be
\cZ_{\Lambda}\simeq \int \lb\prod_f\cA(j_f)\rd (2j_f) \rb \int\rd \mu(X)\, e^{S(\{j_f\},X)+\sum_f 4\pi i u_fj_f }\,,\quad \,u_f\in \Z \, \text{ fixed}\,\,\forall\, f\,.
\label{eq:new_amplitude}
\ee
With a certain choice of face amplitude and a chosen lift of some phase space coordinate 
 to its logarithmic correspondence, the critical deficit angle can take the value $\varepsilon_f=4\pi \Z/\gamma$, similar to the EPRL model. In the special case that $\varepsilon_f=0$ for all $f$, it describes a smooth dS or AdS spacetime because the 4-simplices are all constantly curved with consistent $\Lambda$. 
 
The solution $\varepsilon_f=0$, describing an (A)dS spacetime, only appears as the real critical solution, when we consider the action as a function of real configuration variables. By analytic continuation, we also find a complex critical solution that gives a non-trivial deficit angle hinged by each internal triangle. This means the spinfoam model does not suffer from an ``(A)dSness problem" but allows freedom of intrinsic curvature at the semi-classical regime, just as how the flatness problem in the EPRL model is resolved \cite{Han:2021kll}. 

This paper is organized as follows. In Section \ref{sec:review}, we give a concise review of the spinfoam model with $\Lambda\neq 0$ introduced in \cite{Han:2021tzw}, focusing on constructing the vertex amplitude of the spinfoam model. In Section \ref{sec:edge_amplitude}, we propose a precise form of the edge amplitude that describes the gluing of 4-simplices. We complete the construction in Section \ref{sec:full_amplitude} by further fixing the face amplitudes, which allows us to write the full amplitude for any given 4-complex. We then perform the stationary analysis on the spinfoam amplitude to find the critical solutions. This is done in two parts. Firstly, we derive in Section \ref{sec:stationary_vertex} the real critical solution that describes curved 4-simplices and their gluing. We then focus on the real critical solution to the deficit angle in Section \ref{sec:critical_deficit}. In Section \ref{sec:hormander}, we discuss the complex critical solution and find a non-trivial critical deficit angle. After these general analyses, we give a concrete example with a so-called $\Delta_3$ 4-complex and illustrate the critical behaviour of the corresponding spinfoam amplitude. We conclude and give outlooks in Section \ref{sec:conclusion}. Some details and existing results supporting the analysis are supplied in the Appendices. 

\section{Preliminary: 4D spinfoam with $\Lambda\neq 0$ from boundary Chern-Simons theory}
\label{sec:review}

In this section, we give a concise review of the spinfoam model introduced in \cite{Han:2021tzw} which describes 4D quantum gravity with a non-vanishing cosmological constant $\Lambda$ in Lorentzian signature. For more details, we refer to the original paper \cite{Han:2021tzw} and a more recent one \cite{Han:2023hbe}. 

\subsection{Classical theory}

The starting point is the Plebanski action \cite{Plebanski:1977zz} of the first-order 4D gravity with $\Lambda$ on a 4-ball $\cB_4$: 
\be
S_{\text{Plebanski}}[e,\cA]=-\f12 \int_{\cB_4}\tr\left[\lb \star +\f{1}{\gamma}\rb (e\w e)\w \lb \cF(\cA)+\f{\Lambda}{6}(e\w e)\rb \right]\,,
\label{eq:Plebanski_action}
\ee
where $e$ is the cotetrad one-form valuing in $\sl(2,\bC)$, $\cA$ is an $\sl(2,\bC)$ connection with $\cF(\cA)$ being its curvature two-form, $\star$ is the Hodge star operation and $\gamma$ is the Barbero-Immirzi parameter which takes a real value. 
\eqref{eq:Plebanski_action} can be formulated as a BF action 
\be
S_{\BF}[B,\cA]=-\f12\int_{\cB_4} \tr \left[\lb \star +\f{1}{\gamma}\rb B\w \lb \cF (\cA)+ \f{|\Lambda|}{6} B\rb\right]
\label{eq:LBF_action}
\ee
followed by imposing the {\it simplicity constraint} $B\cong \nu e\w e$ which encodes the sign $\nu:=\sgn(\Lambda)$ of $\Lambda$. 
Consider a path integral of the BF action \eqref{eq:LBF_action}. The (Gaussian) integration in the $B$ field reduces the exponent of the integrand to a second Chern-term, which can be written into two CS actions on the boundary $\partial\cB_4\equiv S^3$ of $\cB_4$. That is,
\be
\int\rd \cA \rd B \, e^{\f{i}{\ell_{\p}^2}S_{\BF}[B,\cA]} 
=\int\rd \cA\, \exp\lb\f{3i}{4\ell_{\p}^2|\Lambda|} \int_{\cB_4}\tr\left[\lb\star+\f{1}{\gamma}\rb\cF(\cA)\w\cF(\cA)\right]\rb
=\int\rd A\rd\Ab \, \exp\lb S_{\CS}[A]+S_{\CS}[\bar{A}] \rb\,,
\label{eq:path_integra_LBF}
\ee
where $\ell_\p=\sqrt{8\pi G\hbar/c^3}$ is the Planck length and the Chern-Simons (CS) action $S_{\CS}[A]$ (\resp $S_{\CS}[\Ab]$) is a function of the self-dual connection $A$ (\resp the anti-self-dual connection $\Ab$) with a complex coupling $t$ (\resp $\tb$). The actions take the form
\be
S_{\CS}[A]=\f{t}{8\pi}\int_{S^3} \tr\left[A\w\rd A+ \f32 A\w A\w A\right]\,,\quad 
S_{\CS}[\Ab]= \f{\tb}{8\pi}\int_{S^3} \tr\left[\Ab\w\rd \Ab+ \f32 \Ab\w \Ab\w \Ab\right]\,,
\label{eq:CS_action}
\ee
where $t=k+is$ and $\tb=k-is$ with $k=\f{12\pi}{\ell_\p^2\gamma|\Lambda|}\in\Z_+\,,s=\gamma k\in\R$.

\begin{figure}[h!]
\centering
\includegraphics[width=0.7\textwidth]{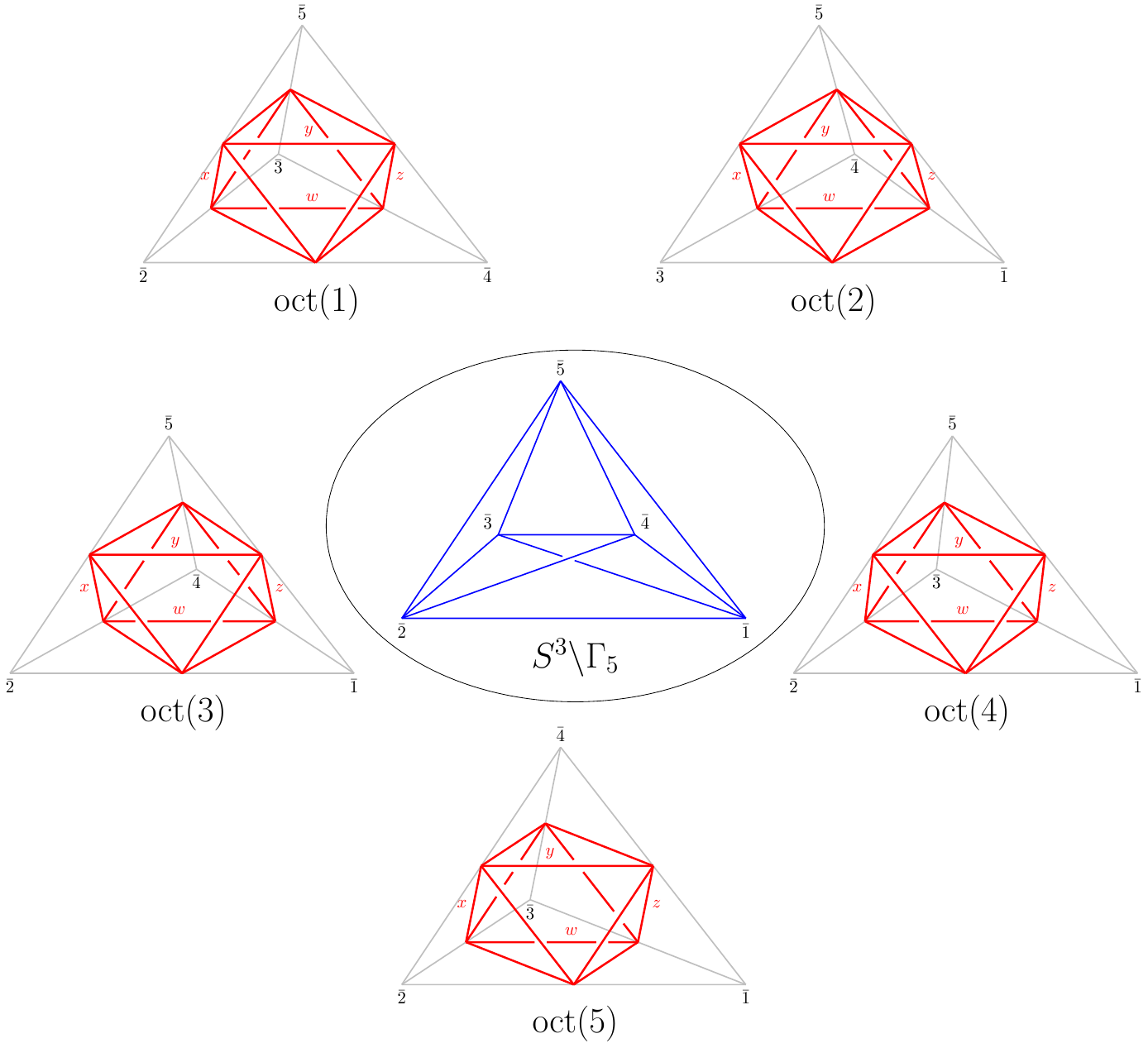}	
\caption{The decomposition of the ideal triangulation $\TSG$ of $\SG$ into 5 ideal octahedra ({\it in red}), each of which can be decomposed into 4 ideal tetrahedra. The cusp boundaries of the ideal octahedra are shrunk to vertices in this figure. 
Numbers $\bar{1},\bar{2},\bar{3},\bar{4},\bar{5}$ with bars denote the 4-holed spheres on $\partial (S^3\backslash\Gamma_3)$. 
In each ideal octahedron, $x, y, z, w$ ({\it labelled in red}) are chosen to form the equator of the octahedron. The same figure appears in \cite{Han:2015gma,Han:2021tzw}. }
\label{fig:triangulation_All}
\end{figure}

Performing the Gaussian integral in $B$ is equivalent to imposing the constraint $\cF=\f{|\Lambda|}{3}B$, so the simplicity constraint now relates the curvature to the cotetrad:
\be
\cF=\f{\Lambda}{3}e\w e\,.
\ee
In order to impose this simplicity constraint later at the quantum level, we introduce {\it defects} on a graph in $S^3$, denoted by $\Gamma_5$ (see the middle graph in blue of fig.\ref{fig:triangulation_All}), which contains 5 nodes and 10 links and can be viewed as the dual graph of the triangulation of $S^3$ -- the boundary of a 4-simplex. 
The defects, carrying the information of the simplicity constraints, generate boundary conditions of the CS theory on the graph-complement $\SG$ and will be quantized to boundary states in the quantum theory. 

\begin{figure}[h!]
\centering
\begin{minipage}{0.2\textwidth}
\centering
\includegraphics[width=0.9\textwidth]{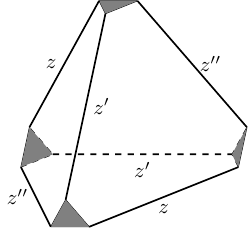}
\subcaption{}
\label{fig:ideal_tetra}
\end{minipage}
\qquad\qquad\qquad\qquad\qquad\qquad
\begin{minipage}{0.25\textwidth}
\centering
\includegraphics[width=0.85\textwidth]{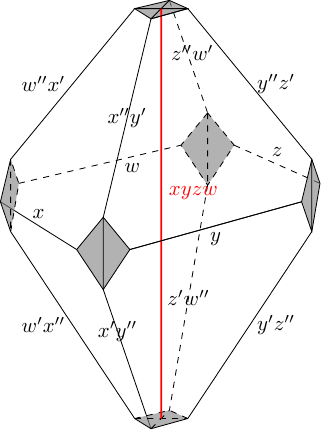}
\subcaption{}
\label{fig:ideal_octa}
\end{minipage}
\caption{{\it (a)} An ideal tetrahedron whose edges are dressed with FGcoordinates $z,z'$ or $z''$. Each pair of opposite edges are dressed with the same coordinate. The cusp boundaries are shown {\it in gray}. {\it (b)} An ideal octahedron. Choose the equator to be edges dressed with $x,y,z,w$. Adding an internal edge ({\it in red}) orthogonal to the equator separates the ideal octahedron into four ideal tetrahedra, each of which is dressed with different copies of coordinates $(x,x',x'')\,,(y,y',y'')\,,(z,z',z'')\,,(w,w',w'')$. For edges shared by different ideal tetrahedra, coordinates are multiplied together.}
\label{fig:ideal_tetra_octa}
\end{figure}

{A set of phase space coordinates on the boundary of $\SG$ can be constructed based on the ideal triangulation of $\SG$, denoted as $\TSG)$, as shown in fig.\ref{fig:triangulation_All}. }
It contains 5 {\it ideal octahedra}, which are octahedra with truncated vertices as shown in fig.\ref{fig:ideal_octa}. Every truncated vertex produces a boundary denoted as a {\it cusp boundary}. By adding an internal edge, an ideal octahedron can be decomposed into 4 {\it ideal tetrahedra}, denoted by $\triangle$, as shown in fig.\ref{fig:ideal_tetra} (see \cite{Han:2021tzw,Han:2023hbe} for more details on such triangulation and see \eg \cite{Gaiotto:2009hg,Dimofte:2011gm,Dimofte:2011ju,Dimofte:2013lba,Dimofte:2014zga,andersen2014complex,Dimofte:2010wxa} for ideal triangulation on manifolds with other topologies). 

On the boundary $\partial\triangle$ of each $\triangle$, the holomorphic CS phase space $\cP_{\partial\triangle}$, which is the moduli space of {\it framed} 
flat connection\footnote{A framed flat connection is a flat connection with a flat section $s$, called the {\it framing flag}, in an associated $\bC\bP^1$ bundle over every cusp boundary, satisfying $(\rd-A)s=0$ \cite{Fock:2003alg,Gaiotto:2009hg,Dimofte:2013lba}.} on $\partial\triangle$, is given by a triple of {\it Fock-Goncharov} (FG) coordinates $(z,z',z'')\in (\bC^*)^{3}$ dressing the edges of $\triangle$ as in fig.\ref{fig:ideal_tetra} subject to a constraint:
\be
\cP_{\partial\triangle}=\{z,z',z''\in\bC^*|zz'z''=-1\}\in(\bC^*)^{ 2}\,.
\label{eq:boundary_phase_space}
\ee
The anti-holomorphic phase space and the corresponding symplectic form are defined in the same way in terms of the complex conjugated FG coordinates $(\bz,\bz',\bz'')$. The constraint as shown in \eqref{eq:boundary_phase_space} eliminates $z'$ and $\bz'$ from the phase space coordinate. The symplectic form of the boundary CS phase space takes the form
\be
\omega_{k,s}=\f{t}{4\pi}\Omega+\f{\tb}{4\pi}\overline{\Omega}\,,\quad \text{with }\,
\Omega=\f{\rd z''}{z''}\w \f{\rd z}{z}\,,\quad 
\overline{\Omega}=\f{\rd \bz''}{\bz''}\w \f{\rd \bz}{\bz}\,,
\label{eq:ABG_symplectic_form_holo}
\ee
which motivates us to take the logarithm of the FG coordinates, each with a randomly chosen but fixed lift: $Z:=\log (z),\,Z':=\log(z'),\,Z'':=\log(z'')$ and similarly for the anti-holomorphic counterparts, so that the canonical pairs have the standard Poisson brackets: $\{Z,Z''\}_\Omega=\{\bZ,\bZ''\}_{\overline{\Omega}}=1$. We reparametrize these logarithmic coordinates in terms of a pair of real variables $(\mu,\nu)\in \R^2$ and a pair of periodic discrete variables $(m,n)\in (\Z/k\Z)^2$ by
\be
Z=\f{2\pi i}{k}\lb-ib\mu-m \rb \,,\quad
Z''=\f{2\pi i}{k}\lb-ib\nu-n \rb\,,\quad
\bZ=\f{2\pi i}{k}\lb-ib^{-1}\mu+m \rb \,,\quad
\bZ''=\f{2\pi i}{k}\lb-ib^{-1}\nu+n \rb \,,
\label{eq:Z_to_mu_m}
\ee
where $b$ is a phase related to the Barbero-Immirzi parameter by $b^2=\f{1-i\gamma}{1+i\gamma}$ with positive real part $\re(b)>0$ and non-zero imaginary part $\im(b)\neq 0$. 

The moduli space of flat connection on $\triangle$, defined by $\cL_\triangle=\{(z,z'')\in \cP_{\partial\triangle}|z^{-1}+z''-1=0\}$ is a Lagrangian submanifold of $\cP_{\partial\triangle}$\footnote{{$x_E=z,z',z''$ can be defined in terms of framing flags parallel transported from the four cusp boundaries of $\triangle$ (see \eqref{eq:FG_from_flag}). One can directly check that, for the case of nilpotent monodromies, the following equations are indeed satisfied.
\be
zz'z''=-1\,,\quad z^{-1}+z''-1=(z'')^{-1}+z'-1=(z')^{-1}+z-1=0\,.
\nn\ee
}}. 
The algebraic curve equation $z^{-1}+z''-1=0$, therefore, restricts the connection on $\triangle$ to be flat. This will play a key role in the critical property of the spinfoam amplitude we analyze in Section \ref{sec:stationary_vertex}. 

\subsection{Quantum theory -- the vertex amplitude}

The new variables $\mu,\nu,m,n$ are quantized into operators and their Poisson brackets are at the same time quantized into commutation relations as follows.
\be
\{\mu,\nu\}_\omega=\{n,m\}_\omega=\f{k}{2\pi}\,,\quad
\{\mu,n\}_\omega=\{\nu,m\}_\omega=0 
\quad\longrightarrow\quad
[\bmu,\bnu]=[\bfn,\bfm]=\f{k}{2\pi i}\,,\quad
[\bmu,\bfn]=[\bnu,\bfm]=0\,.
\ee
The spectra of $\bmu,\bnu$ are real while those of $\bfm,\bfn$ are discrete and bounded to be $\mathbb{Z}/k\mathbb{Z}$. It is then natural to define the kinematical Hilbert space to be $\cH^{\text{kin}}_{k,s}=L^2(\R)\otimes_{\bC} \bC^k$ where $\bC^k$ is a $k$-dimensional vector space.

The building block of the vertex amplitude is provided by the CS partition function on each $\triangle$, which is the {\it quantum dilogarithm function} $\Psi_\triangle(\mu|m)$ 
of the ``position variables'' $(\mu,m)$ of the phase space coordinates 
\be
\Psi_\triangle(\mu|m)=\begin{cases}
	\prod\limits_{j=0}^\infty \f{1-q^{j+1}z^{-1}}{1-\qt^{-j}\zt^{-1}}\,, & \text{ if } |q|>1\\
	\prod\limits_{j=0}^\infty \f{1-\qt^{j+1}\zt^{-1}}{1-q^{-j}z^{-1}}\,, & \text{ if } |q|<1
	\end{cases}\,.
\label{eq:quantum_dilog}
\ee
Here $\mu$ is analytically continued to be a complex variable and $\bz$ is changed to $\zt$ accordingly (as it is no longer the complex conjugate of $z$).
$q$ and $\qt$ encode the CS couplings and play the role of quantum parameters:
\be
q=\exp\lb\f{4\pi i}{t}\rb =\exp\left[\f{2\pi i}{k}(1+b^2)\right]\equiv e^{\hbar}\,,\quad
\qt=\exp\lb\f{4\pi i}{\tb}\rb =\exp\left[\f{2\pi i}{k}(1+b^{-2})\right]\equiv e^{\tilde{\hbar}}\,.
\label{eq:q_qt_def}
\ee
The classical limit is at $\hbar,\tilde{\hbar}\rightarrow 0$ or equivalently $k\rightarrow \infty$. Importantly, $\Psi_\triangle(\mu|m)$ is holomorphic only in the upper half-plane $\im(\mu)>0$ whereas has simple poles at the origin and in the lower half-plane $\im(\mu)\leq0$. More precisely, the poles are located at 
\be
\mu_{\text{pole}}=i b u+ib^{-1}v \quad
\text{with}\,\,u,v\in\Z_-\cup\{0\}\,\,\text{and }\,\,u-v=-m+k\Z\,.
\label{eq:pole_quantum_dilog}
\ee
In order to obtain an absolutely convergent integrals on (Fourier transform of) $\Psi_\triangle(\mu|m)$, which are essential for a finite result of the spinfoam amplitude as we will show later, the integration contour of $\mu$ needs to be shifted to avoid the poles. This was the motivation to introduce imaginary parts $\alpha =\im(\mu)$ and $\beta=\im(\nu)$ to the continuous parameters $\mu$ and $\nu$ respectively in the quantum theory. $(\alpha,\beta)$ are chosen within a region called the {\it positive angle structure} \cite{Dimofte:2014zga} of $\Psi_\triangle$, denoted as $\fP_\triangle$. However, only the real parts $\re(\mu)$ and $\re(\nu)$ are quantized while $\alpha, \beta$ are kept classical. 

Gluing $\triangle$'s reflects symplectic transformations on the phase space coordinates. In addition, a constraint on the FG coordinates is imposed on each internal edge created from gluing $\triangle$'s. Given the (logarithmic) FG coordinates $\{Z_{E}^\triangle,\, \Zt_{E}^\triangle\}_{\triangle\ni {E}}$ on an internal edge ${E}$ from different $\triangle$'s, such a constraint and its quantization take the following form.
\be
C_{E}=\sum_{\triangle \ni {E}} Z_{E}^\triangle =2\pi i \,,\quad
\widetilde{C}_{E}=\sum_{\triangle \ni {E}} \Zt_{E}^\triangle =2\pi i 
\quad\longrightarrow\quad
{\bf C}_{E}=2\pi i+\hbar\,,\quad
\widetilde{\bf C}_{E}=2\pi i+\tilde{\hbar}\,.
\ee

In $\TSG$, internal edges are those added in the ideal octahedra to separate each ideal octahedron into 4 $\triangle$'s (see fig.\ref{fig:ideal_octa}). Consider 4 copies of (logarithmic) FG coordinates $(X,X',X'')\,,(Y,Y',Y'')\,,(Z,Z',Z'')\,,(W,W',W'')$, each for one $\triangle$ in an ideal octahedron. 
The constraints and their quantizations are
\be
\ba{l}
\mu_X+\mu_Y+\mu_Z+\mu_W=0\\
m_X+m_Y+m_Z+m_W=0
\ea 
\quad\longrightarrow\quad
\ba{l}
\bmu_X+\bmu_Y+\bmu_Z+\bmu_W=iQ\\
\bfm_X+\bfm_Y+\bfm_Z+\bfm_W=0
\ea\,,\quad
Q=b+b^{-1}\,,
\label{eq:oct_constraint}
\ee
where $\{\mu_i,m_i\}_{i=X,Y,Z,W}$ are the parameters of different FG coordinate copies defined in the same way as in \eqref{eq:Z_to_mu_m} and $\{\bmu_i,\bfm_i\}_{i=X,Y,Z,W}$ are their quantization respectively. Such constraints allow us to eliminate one set of FG coordinates, say $(W,W',W'')$, by symplectic quotient. As a result, the CS partition function on an ideal octahedron is 
\be
\cZ_{\oct}(x,y,z;\xt,\yt,\zt ) = \prod_{i,j,k,l=0}^{\infty} 
\f{1-q^{i+1}x^{-1}}{1-\qt^{-i}\xt^{-1}}
\f{1-q^{j+1}y^{-1}}{1-\qt^{-j}\yt^{-1}}
\f{1-q^{k+1}z^{-1}}{1-\qt^{-k}\zt^{-1}}
\f{1-q^{l}xyz}{1-\qt^{-l-1}\xt\yt\zt}\,,
\ee
where $(x,y,z;\tilde{x},\tilde{y},\zt )\equiv\exp[(X,Y,Z;\widetilde{X},\widetilde{Y},\widetilde{Z})]$. The positive angle structure $\fP_{\oct}$ of $\cZ_{\oct}$ is different from $\fP_\triangle$ but is proven in \cite{Han:2021tzw} to be a non-empty region. 

Gluing 5 ideal octahedra to form $\TSG$ does not introduce more internal edges but the partition function on $\SG$ is subject to a series of symplectic transformations on the FG coordinates which can be summarized in a symplectic matrix $\bM$ defined as follows.
\be
\mat{c}{\vec{\fQ} \\ \vec{\fP}}
=\bM
\mat{cc}{\vec{\Phi} \\ \vec{\Pi}}
+\mat{c}{i\pi \vec{t}\\0}\,,\quad
\bM=\mat{cc}{\bA & \bB \\ -(\bB^\top)^{-1} & 0}\,,
\label{eq:change_coordinate}
\ee
where $\bA$ and $\bB$ are $15 \times 15$ matrices with integer entries and $\vec{t}$ is a length-15 vector with integer elements. $(\vec{\Phi},\vec{\Pi})^\top$ is a vector of coordinates in the CS phase space $\cP_{\partial(\SG)}\equiv\bigotimes_{i=1}^5 \cP_{\partial(\oct)_i}$ of 5 copies of octahedron boundaries with $\vec{\Phi}=(X_i,Y_i,Z_i)^\top_{i=1,\cdots,5}$ being the position variables and $\vec{\Pi}=(P_{X_i}:=X_i''-W_i'',P_{Y_i}:= Y_i''-W_i'',P_{Z_i}:= Z_i''-W_i'')^\top_{i=1,\cdots,5}$ being the conjugate momenta before the symplectic transformations. $\vec{\fQ}, \vec{\fP}$ are the position and momentum variables after the transformations respectively. 

Parametrizing these coordinates as $\vec{\fQ}=\f{2\pi i}{k}(-ib\vec{\mu}-\vec{m})$ and $\vec{\fP}=\f{2\pi i}{k}(-ib\vec{\nu}-\vec{n})$ with $\vec{\mu},\vec{\nu}\in \bC^{15}\,,\, \vec{m},\vec{n}\in (\Z/k\Z)^{15}$, the resulting partition function is written as \cite{Han:2023hbe}
 \be
\cZ_{\SG}(\vec{\mu}|\vec{m})
=\f{4i}{k^{15}}\sum_{\vec{n}\in(\Z/k\Z)^{15}}\int_{\cC^{\times 15}}\rd^{15}\vec{\nu}\,
{(-1)^{\vec{t}\cdot\vec{n}}}
e^{\f{i\pi}{k}(-\vec{\nu}\cdot \bA\bB^\top\cdot \vec{\nu}+\vec{n}\cdot \bA\bB^\top\cdot \vec{n})}
 e^{\f{2\pi i}{k}\left[-\vec{\nu}\cdot(\vec{\mu}-\f{iQ}{2}\vec{t})+\vec{n}\cdot \vec{m}\right]}\cZ_\times(-\bB^\top\vec{\nu}|-\bB^\top\vec{n})\,,
\label{eq:partition_S3G5}
\ee
where the integration contour $\cC^{\times 15}$ is along $\vec{\beta}:=\im(\vec{\nu})$ which satisfies the positive angle structure $\fP(\SG)$ of $\cZ_{\SG}$. See \cite{Han:2021tzw} for more details on $\fP(\SG)$.  

Different from the fact that elements of $\vec{\Phi}$ and $\vec{\Pi}$ are coordinates on edges of ideal octahedra, elements of $\vec{\fQ}$ and $\vec{\fP}$ are coordinates on annuli, which are the boundaries created by removing the edges of $\Gamma_5$ from $S^3$, and coordinates on 4-holed spheres $\{\cS_a\}_{a=1,\cdots,5}$, which are the boundaries created by removing the 4-valent-nodes of $\Gamma_5$ from $S^3$. 
We denote these coordinates as follows. 
\be
\vec{\fQ}=(\{2L_{ab}\}_{a<b},\{\cX_a\}_{a=1}^{5})\,,\quad 
\vec{\fP}=(\{\cT_{ab}\}_{a<b},\{\cY_a\}_{a=1}^{5})\,,
\label{eq:Q_P}
\ee
where $2L_{ab}$ is called the {\it complex Fenchel-Nielson (FN) length} on the annulus, denoted by $(ab)$ or $(ba)$), connecting $\cS_a$ and $\cS_b$ through holes and its conjugate momentum $\cT_{ab}$ is called the {\it FN twist}\footnote{$L_{ab}$ and $\cT_{ab}$ are (logarithmic) $\SL(2,\bC)$ FN coordinates while $2L_{ab}$ is an $\PSL(2,\bC)$ FN length as $l^2_{ab}:=\exp(2L_{ab})$ can not distinguish $l_{ab}$ and $-l_{ab}$. This is the reason for the introduction of factor 2 for the FN lengths. See \cite{Han:2021tzw,Han:2023hbe} for more discussion.}. 
On the other hand, $\cX_a$ and $\cY_a$ are FG coordinates on $\cS_a$. {Introduce an orientation for each annulus $(ab)$ such that $\cS_a$ is the source and $\cS_b$ is the target if $a<b$ and the opposite if $a>b$. There is a constraint $L_{ba}=-L_{ab}$ on each $(ab)$ due to the gluing of ideal tetrahedra to form ideal octahedra \cite{Han:2021tzw}. }

Simplicity constraints can be separated into first-class type, which we impose strongly, and second-class type, which we impose weakly, at the quantum level as how we treat them in the EPRL model \cite{Engle:2007qf}. 
The first-class simplicity constraints correspond to flat connections on the annuli while the second-class ones correspond to those on the 4-holed spheres. Parametrize each FN length as $2L_{ab}:=\f{2\pi i}{k}\lb -ib\mu_{ab}-m_{ab}\rb$. 
The first-class constraints require that $2L_{ab}\in i\R$ hence $\re(\mu_{ab})=0$. Imposing this quantumly means that we require the partition function, or the quantum state, $\cZ_{S^3\backslash\Gamma_5}(\vec{\mu}|\vec{m})$ to satisfy $\re(\bmu_{ab})\cZ_{S^3\backslash\Gamma_5}(\vec{\mu}|\vec{m})=0$. Such quantum states are those labeled by ``spins'' $j_{ab}:=m_{ab}/2\in\{0,\f12,\cdots,\f{k-1}{2}\}$ dressing the annuli (since $\alpha_{ab}=\im(\mu_{ab})$ is not quantized):
\be
\cZ_{S^3\backslash\Gamma_5}(\{i\alpha_{ab}\}_{a<b}, \{\mu_{a}\}\mid \{j_{ab}\}_{a<b}, \{m_{a}\})\,, 
\label{eq:Z_after_1st_constraints}
\ee
$j_{ab} (a<b)$ encodes the area $\fa_f\equiv\fa_{ab}$ of a curved triangle $f$ on the boundary of a homogeneously curved tetrahedron, which is isomorphic to the moduli space of $\SU(2)$ flat connection on a 4-holed sphere $\cS_a$ \cite{Haggard:2015ima}. For the convenience of some discussion, we also introduce $j_{ba}$ that relates to $m_{ba}$ of $L_{ba}=\frac{2\pi}{k}(-ib\mu_{ba}-m_{ba})$ in the same way. 

Fixing the areas of all the boundary triangles, the (reduced) moduli space of flat connection has a pair of Darboux coordinates, denoted by $(\theta_a,\phi_a)$, as functions of the FG coordinates $\cX_a=\f{2\pi i}{k}\lb -b\mu_a-m_a\rb$ and $\cY_a=\f{2\pi i}{k}\lb -b\nu_a-n_a\rb$. 
The second-class constraints are imposed on these FG coordinates. 
 To impose them weakly, we define a product coherent state $\Psi_{\rho_a}(\re(\mu_a)|m_a):=\psi_{z_a}(\re(\mu_a))\otimes \xi_{(x_a,y_a)}(m_a)$ on each $\cS_a$ living in the Hilbert space $\cH_{\cS_a}= L^2(\R)\otimes_{\bC}\bC^k$ \st $\rho_a=(z_a,x_a,y_a)\in \bC\otimes \bT^2$ is a triple of coherent state labels. The two coherent states defining $\Psi_{\rho_a}$ are
\begin{subequations}
\begin{align}
\psi_{z_a}(\re(\mu))&:= e^{-\sqrt{2}\beta_a\re(z_a)}\lb\f{2}{k}\rb^{1/4} e^{-\f{\pi}{k}\lb\re(\mu)-\f{k}{\pi\sqrt{2}}\re(z_a)\rb^2}e^{-i\sqrt{2}\re(\mu)\im(z_a)}\in L^2(\R)\,,\quad z_a\in\bC\,,
\label{eq:coherent_state_1}\\
\xi_{(x_a,y_a)}(m)&:=\lb\f{2}{k}\rb^{1/4} e^{\f{ikx_ay_a}{4\pi}}\sum_{p_a\in \Z}
e^{-\f{k}{4\pi}\lb \f{2\pi m}{k}-2\pi p_a-x_a \rb^2} e^{\f{ik}{2\pi}y_a\lb \f{2\pi m}{k}-2\pi p_a-x_a\rb}\in \bC^k\,,\quad (x_a,y_a)\in [0,2\pi)\times[0,2\pi)\,,
\label{eq:coherent_state_2}
\end{align}
\label{eq:coherent_state_def}
\end{subequations}
where $\beta_a=\im(\nu_a)$\footnote{The pre-factor $ e^{-\sqrt{2}\beta_a\re(z_a)}$ in defining $\psi_{z_a}(\re(\mu))$ is there for a bounded result of the vertex amplitude \cite{Han:2021tzw} and its contribution is negligible at large-$k$ regime as we will see in the stationary analysis.}.  
The second-class constraints are imposed on the coherent state labels $\{\rho_a\}_a$ and we denote those satisfying the simplicity constraints as $\{\hrho_a=(\zh_a,\hx_a,\yh_a)\}_a$. $\left|\Psi_{\hrho_a}(\re(\mu_a)|m_a)\right|$ peaks at 
\be
\re(\mu_a)=\f{k}{\pi \sqrt{2}}\re(\zh_a)\,,\quad
\Mod(m_a,k)=\f{k}{2\pi}\hx_a\,.
\ee
$\hrho_a$ corresponds to the Darboux coordinates $(\htheta_a,\hphi_a)$ of the moduli space of flat $\SU(2)$ 
flat connection describing the shape of the curved tetrahedron with fixed triangle areas \cite{Han:2023hbe}. 
The range of $(\htheta_a,\hphi_a)$ depends on the 4 spin configurations $\{j_{ab}\}_{b\neq a}$. We denote the range as $\overline{\cM}_{\vec j}$ with continuous $\vec{j}\in [0,k/2)^{\times4}$. 
The second-class simplicity constraints are imposed weakly in the sense that they are satisfied only on the peaks of $\{\left|\Psi_{\hrho_a}\right|\}_a$. 

The vertex amplitude is defined by the inner product of the partition function \eqref{eq:Z_after_1st_constraints} and 5 coherent states $\{\Psi_{\hrho_a}\}_{a=1}^5$. That is,
\be\begin{aligned}
\cA_v(\iota):=&\la\,
\overline{\Psi}_{\hrho_a}\mid\cZ_{S^3\backslash\Gamma_5}\,\ra\\
=&\sum_{\{m_a\}\in (\Z/k\Z)^5}\int_{\R^5} \rd^5\mu_a \, 
\cZ_{S^3\backslash\Gamma_5}(\{i\alpha_{ab}\}_{a<b}, \{\mu_{a}+i\alpha_a\}\mid\{j_{ab}\}_{a<b}, \{m_{a}\})
\prod\limits_{a=1}^5\Psi_{\hrho_a}(\mu_a|m_a)\,,
\label{eq:vertex_amplitude}
\end{aligned}\ee
where $\iota=(\{\alpha_{ab},j_{ab}\}_{a<b}, \{\hrho_a\}_{a=1}^5, \{\alpha_a,\beta_a\}_{a=1}^5)$ with $\beta_a=\im(\nu_a)$,
 and we have specified the imaginary part $\alpha_a$ of the continuous parameter of $\cX_a$ hence $\mu_a\in\R$ in the expression. 
It is proven in \cite{Han:2021tzw} that the vertex amplitude $\cA_v(\iota)$ defined in \eqref{eq:vertex_amplitude} is bounded for any coherent state labels $\{\hrho_a\}_{a=1}^5$. 

The construction of the vertex amplitude of a spinfoam vertex $v$ described above was based on a labelling order $1,2,\cdots,5$ of 4-holed spheres on $\partial(\SG)$ (\eg it relies on $j_{ab}$ with $a<b$ while $\{j_{ba}\}_{a<b}$ are redundant). Such a labelling order dependence can be removed by introducing a sign $\kappa^v_{ab}=\pm 1$ on each annulus that relates to the orientation of the annulus. Then the FN length $L^v_{ab}$ is redefined as  
\be
\kappa^v_{ab} L^v_{ab}= \f{2\pi i}{k}(-ib\mu_{ab}-m_{ab})\,.
\ee
$\kappa_{ab}^v$ satisfies the following properties.
\be
\kappa_{ab}^v=-\kappa_{ba}^v\,,\quad
\kappa_{ac}^v=-\kappa_{bd}^{v'} \,\text{ if annulus $(ac)$ of $v$ connects annulus $(bd)$ of $v'$}\,.
\ee
The previous construction is then a special case when $\kappa^v_{ab}=1$ for $a<b$. 
The introduction of $\kappa^v_{ab}$ is related to the face orientation of a spinfoam 2-complex \cite{Han:2021rjo}.

\section{Gluing $\SG$'s -- the edge amplitude}
\label{sec:edge_amplitude}

\begin{figure}[h!]
    \centering
    \includegraphics[width=0.8\textwidth]{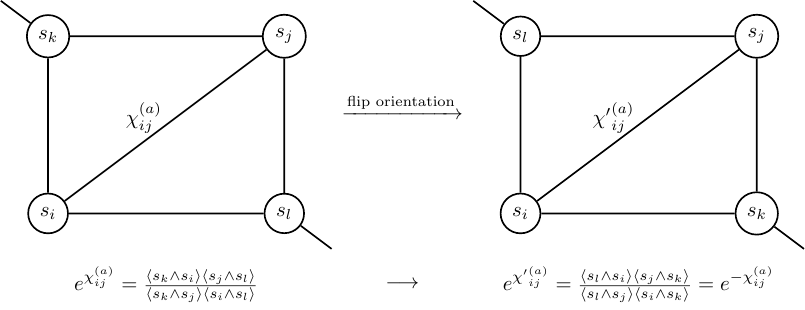}
    \caption{FG coordinate dressing the edge on $T_a$ connecting hole $i$ and $j$ defined from framing flags $\{s_i,s_j, s_k, s_l\}$ parallel transported from holes of $\cS_a$ before and after flipping the orientation of $\cS_a$.}
    \label{fig:FG_flip}
\end{figure}

3-manifolds $\SG$'s are glued through ``eliminating'' boundary 4-holed spheres pairwise. Each gluing contributes an edge amplitude to the spinfoam amplitude. 
Denote the two 4-holed spheres to be glued as $\cS_a^v$ (from spinfoam vertex $v$) and $\cS_b^{v'}$ (from spinfoam vertex $v'$).  The gluing is done by flipping the orientation of one of the 4-holed spheres, say $\cS_b^{v'}$, then identify the holes pairwise. This also automatically identifies the edges of the ideal triangulation $T_a$ of $\cS_a^v$ and that $T_b$ of $\cS_b^{v'}$ pairwise. 
Intuitively, such a gluing should produce constraints semi-classically to the FG coordinates on the glued 4-holed spheres. More precisely, suppose the edge on $T_a$ dressed with $e^{\chi_{ij}^{(a)}}$ is glued to the edge on $T_b$ dressed with $e^{{\chi'}_{i'j'}^{(b)}}$, then 
\be
e^{\chi_{ij}^{(a)}}=e^{-{\chi'}_{i'j'}^{(b)}}\,.
\label{eq:gluing_condition}
\ee
We call this the {\it gluing condition}. 
The minus sign on the ${\it r.h.s.}$ is there since, when one flips the orientation of a 4-holed sphere, the FG coordinate defined in terms of the framing flags (see \eqref{eq:FG_from_flag}) is changed to its inverse as shown in fig.\ref{fig:FG_flip}. 
For each gluing, there are 6 such gluing conditions, each corresponding to an edge of $T_a$ (or an edge of $T_b$). Four of them are given by identifying the 4 spins on the annuli attached to holes of $\cS_a^{v}$ and $\cS_b^{v'}$.
They give the constraints 
\be
{j}_{bd}^{v'}={j}_{ac}^v\equiv j_f
\label{eq:gluing_constraint}
\ee
on the logarithmic FN coordinates {$\{2\kappa_{ac}^vL_{ac}^v=-4\pi i {j}_{ac}^{v} /k\}_c$ and $\{2\kappa_{bd}^{v'}L_{bd}^{v'}=-4\pi i j_{bd}^{v'} /k\}_d$ }
when annuli $(ac)$ and $(bd)$ correspond to the same spinfoam face $f$. 

The remaining two gluing conditions are imposed on the FG coordinates $(e^{\cX_a^v},e^{\cY_a^v})$ on $\cS_a^v$ and those $(e^{\cX_b^{v'}},e^{\cY_b^{v'}})$ on $\cS_b^{v'}$.  
For simplicity, we let the edge dressed with $e^{\cX^v_a}$ be glued to the edge dressed with $e^{-\cY^{v'}_b}$ and let the edge dressed with $e^{-\cY^v_a}$ be glued to the edge dressed with $e^{\cX^{v'}_b}$\footnote{It is not possible that the edge dressed with $e^{\cX_a^v}$ (\resp dressed with $e^{-\cY_a^v}$) is glued to the edge dressed with $e^{\cX_b^{v'}}$ (\resp dressed with $e^{-\cY_b^{v'}}$) as $\cS_a^v$ and $\cS_b^{v'}$ are both oriented outward. See the upper panel of fig.\ref{fig:glue} for an illustration. }. {Such a requirement might give restriction on the topology of the simplicial complex after gluing, but is shown to be possible in some simple examples including the $\Delta_3$-complex as will be shown in Section \ref{sec:delta3}. }
Parameterize the FG coordinates $(e^{\cX_a^v},e^{\cY_a^v},e^{\cX_b^{v'}},e^{\cY_b^{v'}})$ as
\be
e^{\cX_a^v}=e^{\f{2\pi i}{k}(-ib\mu_a^v-m_a^v)}\,,\quad
e^{\cY_a^v}=e^{\f{2\pi i}{k}(-ib\nu_a^v-n_a^v)}\,,\quad
e^{\cX_b^{v'}}=e^{\f{2\pi i}{k}(-ib\mu_b^{v'}-m_b^{v'})}\,,\quad
e^{\cY_b^{v'}}=e^{\f{2\pi i}{k}(-ib\nu_b^{v'}-n_b^{v'})}\,.
\ee

As each 4-holed sphere is coupled with a coherent state, we propose an edge amplitude as a function of coherent state labels that peaks at the gluing condition. 
Denote the coherent state coupled with $\cS_a^{v}$ as $\Psi_{\hrho_a}(\mu_a^v|m_a^v)$ with $\hrho_a=(\zh_a,\hx_a,\yh_a)$ and that coupled with $\cS_b^{v'}$ as $\Psi_{\hrho'_b}(\mu_b^{v'}|m_b^{v'})$ with $\hrho'_b=(\zh'_b,\hx'_b,\yh'_b)$. 
As can be seen from the expression \eqref{eq:coherent_state_def} of the coherent states (and derived in Section \ref{sec:stationary_vertex}), the coherent states $\Psi_{\hrho_a}$ and $\Psi_{\hrho'_b}$ peak at 
\be\ba{lll}
\zh_a = \f{\sqrt{2}\pi}{k}\lb \re(\mu_a^v) - i \re(\nu_a^v)\rb \,,\quad &
\hx_a =\f{2\pi}{k}\Mod(m_a^v,k)\,,\quad &
\yh_a=-\f{2\pi}{k}\Mod(n_a^v,k)\,,\quad\\[0.15cm]
\zh'_b = \f{\sqrt{2}\pi}{k}\lb \re(\mu_b^{v'}) - i \re(\nu_b^{v'})\rb \,,\quad &
\hx'_b =\f{2\pi}{k}\Mod(m_b^{v'},k)\,,\quad &
\yh'_b=-\f{2\pi}{k}\Mod(n_b^{v'},k)\,.
\ea\ee

We define the edge amplitude on a spinfoam edge $e$ corresponding to gluing $\cS_a^v$ and $\cS_b^{v'}$ to be\footnote{The prefactor $\lb\f{k}{4\pi^2}\rb^2$ is inspired by the over-completeness of the coherent state (before imposing the simplicity constraints on $\rho_a$), \ie\be
\lb\f{k}{4\pi^2}\rb^2\int_{\bC\times\bT^2}\rd \rho_a\,\Psi_{\rho_a}(\mu|m)\overline{\Psi}_{\rho_a}(\mu'|m') e^{2\sqrt{2}\beta_a\re(z_a)}=\delta_{\mu,\mu'}\delta_{e^{\f{2\pi i}{k}(m-m')},1}\,.
\nn\ee}
\be
\cA_e(\hrho_a^v,\hrho_b^{v'}|\{j_{ac}^v,j_{bd}^{v'}\}_{c,d})=
\lb\f{k}{4\pi^2}\rb^2\prod_{c,d}\delta_{{j}_{bd}^{v'}, {j}_{ac}^v}
\,\exp\left[kS_e( \hrho_a^v,\hrho_b^{v'} )\right]\,,
\label{eq:edge_amplitude}
\ee
where
\begin{eqnarray}
S_e( \hrho_a^v,\hrho_b^{v'} )&:=&
-\f{1}{4\pi}\lb 2\lb \re(\zh_a)+\im(\zh'_b)\rb^2 +2\lb \re(\zh'_b)+\im(\zh_a)\rb^2+\lb \xh_a+\yh'_b\rb^2+\lb\yh_a+\xh'_b\rb^2\rb \nonumber\\
&& + \f{i}{4\pi} \lb4\im(\zh'_b)\im(\zh_a)  
+ \xh_a\yh_a+\xh'_b\yh'_b+2\yh_a \yh'_b\rb\,.
\label{eq:edge_action}
\end{eqnarray}
Its stationary point is at 
\be
\re(\zh_a)=-\im(\zh'_b)\,,\quad
\im(\zh_a)=-\re(\zh'_b)\,,\quad
\xh_a=-\yh'_b\,,\quad
\yh_a=-\xh'_b\,.
\ee
The imaginary part of the action is there for the existence of non-trivial critical points, which we will derive in Section \ref{sec:stationary_vertex}. 
 
The edge amplitude relates the coherent state labels of the coherent states coupled with $\cS_a^v$ and $\cS_b^{v'}$. 
We denote the integral over the coherent state labels upon the imposition of the simplicity constraints in terms of the Darboux coordinates $(\htheta_a,\hphi_a)$ on $\cS_a$ and those $(\htheta'_b,\hphi'_b)$ on $\cS'_b$ \cite{Han:2023hbe}:
\be
\int_{\overline{\cM}_{\vec{j}_a^{v}}}\rd\hrho_a:=\frac{1}{2Q^2}\int_{\overline{\cM}_{\vec{j}_a^{v}}}\rd\htheta_a\w\rd\hphi_a \,,\quad
\int_{\overline{\cM}_{\vec{j}_b^{v'}}}\rd\hrho_b:=\frac{1}{2Q^2}\int_{\overline{\cM}_{\vec{j}_b^{v'}}}\rd\htheta'_b\w\rd\hphi'_b\,.
\ee
Combining the measures on the coherent state labels, the expression
\be
\sum_{\{j_{ac}^v,j_{bd}^{v'}\}_{c,d}}\int_{\overline{\cM}_{\vec{j}_a^v}}\rd\hrho_a^v\int_{\overline{\cM}_{\vec{j}_b^{v'}}}\rd\hrho_b^{v'}\,
\cA_v(\hrho_a^v,\{j_{ac}^v\}) \cA_e(\hrho_a^v,\hrho_b^{v'}|\{j_{ac}^v,j_{bd}^{v'}\}_{c,d})\cA_{v}(\hrho_b^{v'},\{j_{bd}^{v'}\})
\ee
gives a non-zero and bounded result when $\overline{\cM}_{\vec{j}_a^v}\cap\overline{\cM}_{\vec{j}_b^{v'}}\neq\varnothing$.

\section{The full finite amplitude}
\label{sec:full_amplitude}

The set of gluing constraints \eqref{eq:gluing_constraint} collected from all the gluing processes to obtain the final 3-manifold are not necessarily independent due to the intrinsic symmetry $L_{ab}^v=-L_{ba}^v$ in $\{L_{ab}^v\}$. 
This needs to be taken into account when boundary annuli are glued to form a boundary torus, which corresponds to an internal spinfoam face. In this case, a face amplitude needs to be taken into account. 

Separate the spins $\{j_f\}$ into those for the boundary annuli and those for the boundary tori, or internal spinfoam faces. 
We adopt the face amplitude for each internal spinfoam face proposed in \cite{Han:2023hbe} with an undetermined power $\fp\in\R$: 
 \be
\cA_f(j_f)=[2j_f+1]_\fq^\fp \, {e^{\f{ik}{2\pi}\cF_f(-\f{2\pi i}{k}2j_f)}}
\,,\quad\fp\in \R\,,\quad
j_f=0,\f12,\cdots,\f{k-1}{2}\,,
\label{eq:face}
\ee
where $[n]_\fq:=\f{\fq^n-{\fq}^{-n}}{\fq-\fq^{-1}}\equiv \sin\lb \f{2\pi n}{k} \rb/\sin\lb \f{2\pi}{k}\rb$ is a $\fq$-number with $\fq=e^{2\pi i/k}$ being a root-of-unity depending on the CS level $k$. At $k\rightarrow\infty$ limit, $[2j_f+1]_\fq^\fp\rightarrow(2j_f+1)^\fp$ becomes the face amplitude used in the EPRL model as desired. 
The $\fq$-deformation of this term is due to the following argument. 
The form of the face amplitude is related to the boundary Hilbert space \cite{Bianchi:2010fj}, which we expect to be spanned by spin network states defined from the CS theory. On the other hand, the quantum states of CS theory at level $k$ are described by the quantum group deformation of the gauge group \cite{Alekseev:1994au,Alekseev:1994pa,Alekseev:1995rn}. We, therefore, expect that the spin network states should be $\fq$-deformed, so as the face amplitudes\footnote{The term $[2j_f+1]_\fq^\fp$ in the face amplitude does not affect the semi-classical analysis of the amplitude. In this context, it does no harm to change this factor to any other polynomial of $j_f$ or its $\fq$ deformation.}. 

$\cF_f$ given in \eqref{eq:face} is a real quadratic function in $2L_{f}$ defined as
\be
\cF_f(2L_{f}):= a_f \lb2L_f\rb^2 +i\pi b_f\cdot 2L_f+c_f\,,\quad
a_f,\,b_f,\,c_f\in\R\,.
\label{eq:cF}
\ee
The coefficients are undetermined at this stage, but we will come back to them when we consider the critical equation in Section \ref{sec:critical_deficit}. {The addition of $\exp \lb \f{ik}{2\pi}\cF_f \rb$ in \eqref{eq:face} 
is related to the fact that the FN twist, denoted as $T_f$, conjugate to $2L_f$ and associated to the B-cycle (along the longitude) of a boundary torus is not necessarily only the linear combination of the conjugate momenta $\{\cT^v_f\}_v$ but may also contain a term linear in $2L_f$ (see the discussion below \eqref{eq:tau_chi}). As we will see in Section \ref{sec:critical_deficit}, adding this term and carefully choosing the values of $a_f$ and $b_f$ can reproduce the information of $T_f$ in the semi-classical regime.} 
On the other hand, the addition of this term only changes the phase of the CS wave functions, which is not important in the CS theory as they are not seen when constructing physical observables. However, it plays a role in the spinfoam amplitude when internal spins are summed over. 
It is particularly important for analyzing the critical point of the action \wrt the derivative of $2L_f$, which we will see in Section \ref{sec:critical_deficit}. 

\bigskip

In summary, the spinfoam amplitude for a {spinfoam 2-complex} consisting of $V$ spinfoam vertices, $E_{\In}$ internal spinfoam edges and $F_{\In}$ internal spinfoam faces takes the form 
\be
\cZ_{\vec{\hrho}_\partial}(\vec{\alpha}|\vec{j}_b)
=\sum_{j_f=0}^{(k-1)/2}
\int_{\overline{\cM}_{\vec{j}^v_a}}\rd \hrho^{v\in e}_a\int_{\overline{\cM}_{\vec{j}^{v'}_b}}\rd \hrho^{v'\in e}_b \,
\left[\prod_{f=1}^{F_{\In}}\mathcal{A}_f(2j_f)\right]
\left[\prod_{e=1}^{E_{\In}}
\cA_e(\hrho_a^{v\in e},\hrho_b^{v'\in e}|\{j_{ac}^{v\in e},j_{bd}^{v'\in e}\}_{c,d})\right]
\left[\prod_{v=1}^V\cA_{v}(\vec{\alpha}^v,\vec{j}^v,\vec{\hrho}^v)\right],
\label{eq:spinfoam_amplitude}
\ee
where $v\in e$ denotes that $v$ is at the (source or target) end of $e$, $\vec{\alpha}$ contains all the positive angles, $\vec{\hrho}_\partial$ contains all the coherent state labels on the boundary, the summations in $j_f$ are for all the internal spinfoam faces and the integrations over coherent state labels are for all the internal spinfoam edges. 

Now that the vertex amplitudes, edge amplitudes and face amplitudes are all bounded by the above construction, and that the integrations over the coherent state labels are over compact domains, 
the spinfoam amplitude defined in \eqref{eq:spinfoam_amplitude} 
for {\it any} spinfoam 2-complex is {\it finite} given finite boundary spins $\vec{j}_b$ and finite CS level $k$.

\section{Stationary analysis for vertex amplitude and curved Regge action}
\label{sec:stationary_vertex}

In order to analyze the critical point of the spinfoam amplitude \eqref{eq:spinfoam_amplitude}, we look into its large-$k$ regime, where critical equations of the action can be found. 
For that purpose, we convert the parameters $\{\mu_I,\nu_I,m_I,n_I\}$ of all the FG and FN coordinates in $\{\cA_v\}$ and $\{\cA_f\}$ into the coordinates $\{\fQ_I,\fP_I\}$ that do not scale with $k$ by the relations 
\begin{subequations}
\begin{align}
\mu_I=\f{kb}{2\pi(b^2+1)}\lb\fQ_I+\tfQ_I\rb\,,\quad &
m_I=\f{ik}{2\pi(b^2+1)}\lb\fQ_I-b^2\tfQ_I\rb\,,
\label{eq:m_mu-FN_FG}\\
\nu_I=\f{kb}{2\pi(b^2+1)}\lb\fP_I+\tfP_I\rb\,,\quad &
n_I=\f{ik}{2\pi(b^2+1)}\lb\fP_I-b^2\tfP_I\rb\,.
\label{eq:n_nu-FN_FG}
\end{align}
\label{eq:mn_munu-FN_FG}
\end{subequations}

It is shown in \cite{Han:2021tzw} that the vertex amplitude \eqref{eq:vertex_amplitude}, at the large-$k$ regime, can be written as
\be
\cA_v(\vec{\alpha}^v,\vec{j}^v,\vec{\hrho}^v)\xrightarrow{k\rightarrow\infty} \cN 
\sum_{\vec{p}^v\in\Z^{15}}\sum_{\vec{u}^v\in\Z^5}
\int_{\cC^{\times40}_{M_v}}\rd M_v\,
\exp\left[ k
S^{v}_{\vec{p}^v,\vec{u}^v,\vec{\hrho}^v}
(\vec{\fP}^v,\vec{\tfP}^v,\vec{\fQ}^v,\vec{\tfQ}^v)
\right]\left[1+O(1/k)\right]\,,
\label{eq:vertex_large_k}
\ee
where the overall constant is $\cN=\f{16\sqrt{2}}{(2\pi)^{40}Q^{20}}k^{45/2}$ and the measure contains the contour integration over all the momenta $\vec{\fP}^v$ and $\vec{\tfP}^v$ and the FG positions $\{\cX_a^v,\widetilde{\cX}^v_a\}_{a=1}^5$ on the 4-holed spheres on each $S^3\backslash\Gamma_5$. Explicitly,
\be
\int_{\cC^{\times40}_{M_v}}\rd M_v:=
\int_{\cC^{\times 30}_{\fP^v\times\tfP^v}}
\bigwedge_{I=1}^{15}\left(-i\, \mathrm{d} \mathfrak{P}^v_{I} \wedge \mathrm{d} \widetilde{\mathfrak{P}}^v_{I}\right)
\int_{\cC^{\times10}_{\cX^v_{a}\times\widetilde{\cX}^v_{a}}}
\bigwedge_{a=1}^{5}\lb-i\, \mathrm{d} {\cX}^v_{a} \wedge \mathrm{d} \widetilde{\cX}^v_{a}\rb\,.
\label{eq:dM}
\ee
The action in \eqref{eq:vertex_large_k} can be separated into several parts as follows.  
\begin{multline}
S^{v}_{\vec{p}^v,\vec{u}^v,\vec{\hrho}^v}
=S^v_0(\vec{\fP}^v,\vec{\tfP}^v,\vec{\fQ}^v,\vec{\tfQ}^v)
+S_1^v(-\bB_v^\top\cdot \vec{\fP}^v)
+\tS^v_1(-\bB_v^\top \cdot\vec{\tfP}^v)
-\f{1}{b^2+1}\vec{p}^v\cdot(\vec{\fP}^v-b^2\vec{\tfP}^v)\\
+\sum\limits_{a=1}^5
\left[S_{\zh^v_{a}}(\cX^v_{a},\widetilde{\cX}^v_{a})
+S_{(\xh^v_{a},\yh^v_{a})}(\cX^v_{a},\widetilde{\cX}^v_{a})
-\f{1}{b^2+1}u^v_{a}\lb\cX^v_{a}-b^2\widetilde{\cX}^v_{a}\rb\right]\,.
\label{eq:effective_action_all}
\end{multline}
The vectors $\vec{p}^v\in\Z^{15}$ and $\vec{u}^v=(u_1^v,\cdots,u_5^v)^\top\in\Z^5$ come from the Poisson resummations of $\vec{n}^v=\f{ik}{2\pi(b^2+1)}\lb\vec{\fP}^v-b^2\vec{\tfP}^v\rb$ (recall the expression \eqref{eq:partition_S3G5}) and $m^v_a=\f{ik}{2\pi(b^2+1)}\lb \cX_a^v-b^2\widetilde{\cX}^v_a \rb$ respectively.  
Neglecting the subleading contributions at large $k$, the first three terms on the {\it r.h.s.} of \eqref{eq:effective_action_all} are explicitly \cite{Han:2021tzw} 
\begin{subequations}
\begin{align}
S_0^v(\vec{\fP}^v,\vec{\tfP}^v,\vec{\fQ}^v,\vec{\tfQ}^v)
=&
-\f{i}{4\pi(b^2+1)}\left[\vec{\fP}^v\cdot\lb\bA\bB^\top\cdot\vec{\fP}^v+2\vec{\fQ}^v \rb 
+b^2\vec{\tfP}^v\cdot\lb\bA\bB^\top\cdot\vec{\tfP}^v
+2\vec{\tfQ}^v \rb\right]
-\f{\vec{t}\cdot \lb\vec{\fP}^v-b^2\vec{\tfP}^v \rb}{2(b^2+1)}\,,
\label{eq:S0}\\
S_1^v(-\bB^\top \cdot\vec{\fP}^v)
=&-\f{i}{2\pi(b^2+1)}\sum_{i=1}^5
\left[\Li_2(e^{-X^v_i})+\Li_2(e^{-Y^v_i})+\Li_2(e^{-Z^v_i})+\Li_2(e^{-W^v_i}) \right]\,,
\label{eq:S1}\\
\tS^v_1(-\bB^\top \cdot\vec{\tfP}^v)
=&-\f{i}{2\pi(b^{-2}+1)}\sum_{i=1}^5
\left[\Li_2(e^{-\Xt^v_a})+\Li_2(e^{-\Yt^v_i})+\Li_2(e^{-\Zt^v_i})+\Li_2(e^{-\Wt^v_i}) \right]\,,
\label{eq:S1t}
\end{align}
\label{eq:S0-S1t}
\end{subequations}
where $-\bB^\top\cdot \vec{\fP}^v=(X^v_i,Y^v_i,Z^v_i)_{i=1}^5$ with subscript $i$ denoting the octahedron $\Oct(i)$ on the $\SG$ corresponding to $v$. Similarly for the tilde sectors. 
The first two actions in the square bracket of \eqref{eq:effective_action_all} correspond to the coherent states \eqref{eq:coherent_state_1} and \eqref{eq:coherent_state_2} respectively and, neglecting subleading contributions, they read
\begin{subequations}
\begin{align}
S_{\zh^v_{a}}(\cX^v_{a},\widetilde{\cX}^v_{a})&=
-\f{ b}{2\pi(b^2+1)}\lb\cX^v_{a}+\widetilde{\cX}^v_a\rb
\left[\f{b\lb\cX^v_{a}+\widetilde{\cX}^v_{a}\rb}{2(b^2+1)}-\sqrt{2}\hat{\bar{z}}^v_{a}\right]-\f{1}{2\pi}\re(\zh^v_{a})^2\,,
\label{eq:coherent_QQt_1}\\
S_{(\xh^v_{a},\yh^v_{a})}(\cX^v_{a},\widetilde{\cX}^v_{a})
&=- \f{i\xh^v_{a}\yh^v_{a}}{4\pi}-\f{1}{4\pi}\left[\f{i\lb\cX^v_{a}-b^2\widetilde{\cX}^v_{a}\rb}{b^2+1}-\xh^v_{a}\right]^2-\f{1}{2\pi}\f{\lb\cX^v_{a}-b^2\widetilde{\cX}^v_{a}\rb  \yh^v_{a}}{b^2+1} 
\,.
\label{eq:coherent_QQt_2}
\end{align}
\label{eq:coherent_QQt}
\end{subequations} 
The action of the full amplitude includes the actions of all the spinfoam vertices and the exponents of the phase factors in all the face amplitudes (when expressing {$2j_f=m_f=\f{ik}{2\pi}\cdot 2L_f$} with $2L_f\equiv\kappa_{ab}2L_{ab}$ for annulus $(ab)$), \ie 
\be
S=\sum_{v=1}^V S^v_{\vec{p}^v,\vec{u}^v,\vec{\hrho}^v}+ \sum_{e=1}^{E_{\In}} S_e(\hrho^{v\in e}_a,{\hrho}^{v'\in e}_b)
+\sum_{f=1}^{F_{\In}} \lb \f{i}{2\pi} \cF_f(2L_f)-2u_fL_f\rb\,,
\label{eq:effective_action}
\ee
where $u_f\in \Z$ comes from the Poisson resummation of $m_f\equiv \f{ik}{\pi}L_f$:
\be
{\sum_{m_f \in \mathbb{Z} / k \mathbb{Z}} \cdots =\frac{k}{2\pi}\sum_{u_f\in\mathbb{Z}}\int_{-\delta/k}^{2 \pi-\delta/k} \mathrm{d} (i2L_f)e^{ -2k u_f L_f}\cdots\,.}
\label{eq:poisson_resum_Qf}
\ee

We now look for the critical points of the action \eqref{eq:effective_action} \wrt the integration variables in the measure \eqref{eq:dM}, which are independent in $S^v\equiv S^v_{\vec{p}^v,\vec{u}^v,\vec{\hrho}^v}$ of different $v$'s. 
(In contrast, $2L_f$'s are entangled among spinfoam vertices and we postpone the analysis on the critical points \wrt them to the next section.)
The critical equations are
\begin{subequations}
\begin{align}
\f{\partial S^v}{\partial \fP^v_{I}}
&=\f{\partial S^v}{\partial \tfP^v_{I}}
=0\,,\quad \forall I=1,\cdots,15\,,
\label{eq:critical_equations_P}\\
\f{\partial S^v}{\partial \cX^v_{a}}
&=\f{\partial S^v}{\partial \widetilde{\cX}^v_a}
=0\,,\quad  \forall a=1,\cdots,5\,.
\label{eq:critical_equations_cX}
\end{align}
\label{eq:critical_equations}
\end{subequations}
\eqref{eq:critical_equations_P} are the reformulations of the algebraic curve equations 
\be
e^{-\vec{\Phi}^v}+e^{\vec{\Pi}^v}-\vec{1}=0\,,\quad 
e^{-\vec{\widetilde{\Phi}}^v}+e^{\vec{\widetilde{\Pi}}^v}-\vec{1}=0
\label{eq:A-poly_oct}
\ee
in terms of the new position and momentum variables $(\vec{\fQ}^v,\vec{\fP}^v)$ and $(\vec{\tfQ}^v,\vec{\tfP}^v)$. Here, $\vec{1}$ is a length-15 constant vector with elements 1. We refer to \cite{Han:2021tzw} for detailed derivation. See also \cite{Han:2023hbe}. 
The solutions to \eqref{eq:A-poly_oct} describe the moduli space $\cL_{\SG}$ of $\SL(2,\bC)$ flat connection on $\SG$ corresponding to $v$, which is a Lagrangian submanifold of the moduli space $\cP_{\partial(\SG)}$ of $\SL(2,\bC)$  flat connection on $\partial(\SG)$ spanned by $(\vec{\Phi}^v,\vec{\Pi}^v)$ and $(\vec{\widetilde{\Phi}}^v,\vec{\widetilde{\Pi}}^v)$. At the same time, the solutions fix the integer vector $\vec{p}^v$ by the lifts of the (exponential) FG coordinates to their logarithmic counterparts. 

On the other hand, the solutions to \eqref{eq:critical_equations_cX} give the expectation values of the FG coordinates under the coherent state basis, which are determined by the coherent state labels\footnote{We use the coherent states in \cite{Han:2023hbe} (up to a pre-factor), which is the complex conjugate of those in the original paper \cite{Han:2021tzw}, hence the critical solutions to $\cX_a^v$ and $\widetilde{\cX}_a^v$ matches the ones in the former paper.}:
\be
\re(\mu^v_{a})=\f{k}{\sqrt{2}\pi}\re(\zh^v_{a})\,,\quad
\re(\nu^v_{a})=-\f{k}{\sqrt{2}\pi}\im(\zh^v_{a})\,,\quad
m^v_{a}=\f{k}{2\pi}\xh^v_{a}\,,\quad
n^v_{a}=-\f{k}{2\pi}\yh^v_{a}\,,\quad
\forall \, a=1,\cdots, 5\,.
\label{eq:critical_solution_Q}
\ee

Since the {\it constrained} coherent state labels $(\zh_a^v,\hx_a^v,\yh_a^v)$ for a 4-holed sphere $\cS_a$ describe the shape of the tetrahedron isomorphic to $\cS_a$ given areas of the boundary triangles fixed by $\{j_{ab}\}_b$, the critical points of the vertex amplitude describe the moduli space $\cM_\Flat(\SG,\SU(2))$ of $\SU(2)$ flat connection. Ref.\cite{Haggard:2014xoa} has revealed that there is an isomorphism between the fundamental group $\pi_1(\SG)$ of $\SG$ and the fundamental group $\pi(\text{4-simplex})$ of the 4-simplex isomorphic to the bulk $\cB_4$ of $S^3$. 
We, therefore, conclude that the peaks of the vertex amplitude describe the curve geometry of the 4-simplex dual to the spinfoam vertex. 

Such a geometrical interpretation can be made exact thanks to the geometrical interpretation of the FN lengths and the FN twists as proven in \cite{Haggard:2014xoa} (see also \cite{Haggard:2015nat,Haggard:2015sl,Han:2015gma}), which we now use. 
Briefly speaking, the FN length $2L_{ab} (a<b)$ measures the area $\fa_{ab}$ of the curved triangle $f_{ab}$ shared by tetrahedron $a$ and $b$ on the boundary of the 4-simplex while the conjugate momentum $\cT_{ab}$ relates the dihedral angle $\Theta_{ab}$ between $a$ and $b$ around $f_{ab}$. We sketch the derivation in Appendix \ref{sec:app_geo}. 

Lastly, we also need to consider the critical points \wrt the coherent state labels $\hrho_a=(\zh_a,\hx_a,\yh_a)$ and $\hrho'_b=(\zh'_b,\hx'_b,\yh'_b)$ associated to the internal spinfoam edges corresponding to gluing $\cS_a$ and $\cS'_b$. Only two real degrees of freedom, captured by $(\htheta_a,\hphi_a)$ (\resp $(\htheta'_b,\hphi'_b)$), are independent in  the four real parameters $(\zh_a,\hx_a,\yh_a)$ (\resp $(\zh'_b,\hx'_b,\yh'_b)$) due to the imposition of the simplicity constraints. 
Indeed, the parts of action $S$ that have dependence on $(\htheta_a,\hphi_a)$ and $(\htheta'_b,\hphi'_b)$ respectively are ({\it r.f.} \eqref{eq:edge_action} and \eqref{eq:coherent_QQt})
\be
S_{\hrho_a} :=S_{\zh^{s(e)}_a}+S_{(\hx_a^{s(e)},\yh_a^{s(e)})}+S_e( \hrho_a^v,\hrho_b^{v'} )\,,\quad
S_{\hrho'_b} :=S_{{\zh}_b^{'t(e)}}+S_{({\hx}_a^{'t(e)},{\yh}_a^{'t(e)})}+S_e( \hrho_a^v,\hrho_b^{v'} )\,.
\ee
Therefore, the critical equations are
\be
 \f{\partial S_{\hrho_a}}{\partial \htheta_a}= \f{\partial S_{\hrho_a}}{\partial \hphi_a}
 =\f{\partial S_{\hrho'_b}}{\partial \htheta'_b} 
 =\f{\partial S_{\hrho'_b}}{\partial \hphi'_b}=0\,.
\label{eq:eom_coherent_special}
\ee
Explicitly, $\partial S_{\hrho_a}/\partial \htheta_a$ is calculated by
\be
\f{\partial S_{\hrho_a}}{\partial \htheta_a}
=\f{\partial S_{\hrho_a}}{\partial \re(\zh_a)}\f{\partial \re(\zh_a)}{\partial \htheta_a} 
+ \f{\partial S_{\hrho_a}}{\partial \im(\zh_a)}\f{\partial \im(\zh_a)}{\partial \htheta_a}
+\f{\partial S_{\hrho_a}}{\partial \hx_a}\f{\partial \hx_a}{\partial \htheta_a}
+\f{\partial S_{\hrho_a}}{\partial \yh_a}\f{\partial \yh_a}{\partial \htheta_a}\,,
\ee
and similarly for the other three partial derivatives in \eqref{eq:eom_coherent_special}. 
A further subset of solutions to \eqref{eq:eom_coherent_special} are then given by solutions to 
\begin{subequations}
\begin{align}
\f{\partial  S_{\hrho_a}}{\partial \re(\zh_a)}
=\f{\partial  S_{\hrho_a}}{\partial \im(\zh_a)}
=\f{\partial  S_{\hrho_a}}{\partial \hx_a}
=\f{\partial  S_{\hrho_a}}{\partial \yh_a}
&=0\,,
\label{eq:eom_coherent_final_1}\\
\f{\partial  S_{\hrho'_b}}{\partial \re(\zh'_b)}
=\f{\partial S_{\hrho'_b}}{\partial \im(\zh'_b)}
=\f{\partial  S_{\hrho'_b}}{\partial \hx'_b}
=\f{\partial  S_{\hrho'_b}}{\partial \yh'_b}
&=0\,.
\label{eq:eom_coherent_final_2}
\end{align}
\label{eq:eom_coherent_final}
\end{subequations}
The relevant actions for the first two equations in \eqref{eq:eom_coherent_final_1} and \eqref{eq:eom_coherent_final_2} respectively are 
\begin{subequations}
\begin{align}
S_e^{\zh_a}&:=-\f{1}{2\pi} \lb\f{\pi\sqrt{2}}{k}\re(\mu_a) -\re(\zh_a) \rb^2 -\f{1}{2\pi}\lb \re(\zh_a)+\im(\zh'_b)\rb^2 +i \lb \f{1}{\pi}\im(\zh'_b)+\f{\sqrt{2}}{k}\re(\mu_a)\rb\im(\zh_a)  \,,\\
S_e^{\zh'_b}&:=-\f{1}{2\pi} \lb\f{\pi\sqrt{2}}{k}\re(\mu'_b) -\re(\zh'_b) \rb^2 -\f{1}{2\pi}\lb \re(\zh'_b)+\im(\zh_a)\rb^2 +i \lb \f{1}{\pi}\im(\zh_a)+\f{\sqrt{2}}{k}\re(\mu'_b)\rb\im(\zh'_b)  \,,
\end{align}
\end{subequations}
where $\mu_a$ (\resp $\mu'_b$) is the parameter of $\cX^v_a$ (\resp $\cX_b^{v'}$) on $\cS_a$ (\resp $\cS'_b$) that enters the variables of $\psi_{\zh_a}$ (\resp $\psi_{\zh'_b}$) coupled to $\cS_a$ (\resp $\cS'_b$). On the other hand, the relevant actions for the last two equations in \eqref{eq:eom_coherent_final_1} and \eqref{eq:eom_coherent_final_2} respectively are
\begin{subequations}
\begin{align}
S_e^{(\hx_a,\yh_a)} &:=  -\f{1}{4\pi}\lb \lb \f{2\pi m_a}{k} - 2\pi p_a-\hx_a \rb^2 
+\lb \xh_a+\yh'_b\rb^2+\lb\yh_a+\xh'_b\rb^2\rb+ i\f{1}{2\pi}\yh_a\lb \f{2\pi m_a}{k}-2\pi p_a+ \yh'_b\rb\,,\\
S_e^{(\hx'_b,\yh'_b)} &:= -\f{1}{4\pi}\lb\lb \f{2\pi m'_b}{k} - 2\pi p_b+\hx'_b \rb^2+  \lb \xh_a+\yh'_b\rb^2+\lb\yh_a+\xh'_b\rb^2\rb +i\f{1}{2\pi}\yh'_b\lb \f{2\pi m'_b}{k}-2\pi p_b+ \yh_a\rb\,,
\end{align}
\end{subequations}
where $m_a$ (\resp $m'_b$) is the parameter of of $\cX^v_a$ (\resp $\cX_b^{v'}$) on $\cS_a$ (\resp $\cS'_b$) that enters the variables of $\xi_{(\hx_a,\yh_a)}$ (\resp $\xi_{(\hx'_b,\yh'_b)}$) coupled to $\cS_a$ (\resp $\cS'_b$), and $p_a,p_b\in\Z$.
Combining \eqref{eq:critical_solution_Q}, the critical solutions to \eqref{eq:eom_coherent_final} are then
\be
\left|\ba{l}
\re(\mu_a)=\re(\nu'_b)=\f{k}{\pi \sqrt{2}}\re(\zh_a)=-\f{k}{\pi \sqrt{2}}\im(\zh'_b)\\[0.15cm]
\re(\mu'_b)=\re(\nu_a)=\f{k}{\pi \sqrt{2}}\re(\zh'_b)=-\f{k}{\pi \sqrt{2}}\im(\zh_a)
\ea\right.\,,\quad
\left|\ba{l}
x_a=-y_b'=2\pi\lb \f{m_a}{k}-p_a\rb =\f{2\pi}{k}\Mod(n_b',k)\\[0.15cm]
x'_b=-y_a=2\pi\lb \f{m'_b}{k}-p_b\rb=\f{2\pi}{k}\Mod(n_a,k)
\ea\right.\,.
\label{eq:sol_coherent_simple}
\ee
When $m_a,n_a,m'_b,n'_b$ are restricted to $[0,k)$, which fix the lifts of the corresponding exponential FG coordinates, the two solutions on the right give $p_a=p_b=0$ and $m_a=n'_b,\, m'_b=n_a$. Therefore, at the peak of the amplitude, the desired gluing condition $\cX_a=\cY'_b,\, \cY_a=\cX'_b$ ({\it r.f. \eqref{eq:gluing_condition}}) is realized. 

Although the the full set of solutions to \eqref{eq:eom_coherent_special} is complex as the explicit expressions of the functions $\htheta_a(\hrho_a)$ and $\hphi_a(\hrho_a)$ (as well as $\htheta'_b(\hrho'_b)$ and $\hphi'_b(\hrho'_b)$) take more complicated forms, only the simple and special solution \eqref{eq:sol_coherent_simple} gives dominant contribution to the amplitude while the contributions from other solutions are negligible. 
This is because the real part of the action 
\be
S_{\hrho_a,\hrho'_b} := S_{\zh^{s(e)}_a}+S_{(\hx_a^{s(e)},\yh_a^{s(e)})}+S_{{\zh}_b^{'t(e)}}+S_{({\hx}_a^{'t(e)},{\yh}_a^{'t(e)})}+S_e( \hrho_a^v,\hrho_b^{v'})
\ee
is zero only at the solution \eqref{eq:sol_coherent_simple} while negative at any other solution, if exists. This means (the absolute value of) the amplitude exponentially decays at large $k$ unless the solution \eqref{eq:sol_coherent_simple} is reached. 
At the dominant solution to all the critical equations (except for the internal spins, which we leave for the next section to look into), 
the FG coordinates $\chi_{ij}^{(a)}$ and ${\chi'}_{kl}^{(b)}$ on each pair of glued edges satisfy
\be
\chi_{ij}^{(a)} = -{\chi'}_{kl}^{(b)}
\ee
as desired.

\section{Stationary analysis for spins and the critical deficit angle}
\label{sec:critical_deficit}

In the previous section, we have only considered the critical equations \wrt the momenta $\vec{\fP}^v$ and $\vec{\tfP}^v$ and positions on 4-holed spheres $\{\cX_a,\widetilde{\cX}_a\}$ which are independent from different vertex amplitudes, and coherent state labels that are relevant only to neighbouring spinfoam vertices. In this section, we further consider the critical equations of the total action \eqref{eq:effective_action} \wrt the internal FN lengths $2L_f$ and discuss their geometrical interpretations. This section and the next contribute to the main result of the current paper. The geometrical interpretations of the FN coordinates described in Appendix \ref{sec:app_geo} will be used to obtain the final result. 

An FN length $2L_{ab}$ becomes internal when the annulus $(ab)\equiv f$ is glued from both ends to become a torus. In the triangulation language, it corresponds to an internal triangle shared by tetrahedra from different 4-simplices. On this boundary torus, a pair of Darboux coordinates is provided by $\lambda^2_f=e^{2L_{f}}$, which is an A-cycle (along the meridian) holonomy eigenvalue, and $\tau_{f}=e^{T_f}$, which is a B-cycle (along the longitude) holonomy eigenvalue. The B-cycle is particularly chosen to be the one that corresponds to the dihedral angles hinged by the triangle dual to $f$ in all tetrahedra sharing this triangle. More precisely\footnote{{To derive \eqref{eq:FN_twist_deficit}, we use the definition \eqref{eq:FN-twist_def} of the (exponential) FN twist, which is also valid for a torus. In the torus case, $\cS_a=\cS_b$, and we can choose $s_{ac}(\fp_a)=s_{be}(\fp_b)$ and $s_{ad}(\fp_a)=s_{bf}(\fp_b)$. In contrast, $s_{ac}(\fp)=G_f^{-1}s_{be}(\fp)$ and $s_{ad}(\fp)=G_f^{-1}s_{bf}(\fp)$ as these framing flags are related by parallel transport with holonomy $G_f\in\SL(2,\bC)$ along the B-cycle of the torus. Along the same calculation as in \eqref{eq:s2xi} -- \eqref{eq:tau_chi}, $e^{-\f12\nu\varepsilon^{(s)}_f + i\theta_f }$. We then use the fact that $\theta_f=0$ 
derived in \eqref{eq:zero_deficit}. (We do not consider the time non-oriented case hence the value $\theta_f=\pi$ is abandoned here.)}}, 
\be
\tau_f=  e^{-\f12\nu\varepsilon^{(s)}_f}\,,\quad
\varepsilon^{(s)}_f=\sum_{v\in f} s_v\Theta_f^v\,,\quad s_v=\sgn(V_4^v)\,,
\label{eq:FN_twist_deficit}
\ee
$\varepsilon_f^{(s)}$ is called the {\it dressed} deficit angle. 
Consider the logarithmic FN twist 
\be
T_f =- \f12 \nu \varepsilon^{(s)}_f +2\pi i N_f\,,\quad
\widetilde{T}_f = - \f12\nu \varepsilon^{(s)}_f -2\pi i N_f\,,\quad\text{with }\, N_f\in\Z \,.
\ee
where $N_f$ specifies the lift from $\tau_f$ to $T_f$. Denote the (signed) momenta conjugate to $\kappa_{ab}2L_{ab}$ and $\kappa_{ab}2\widetilde{L}_{ab}\equiv -\kappa_{ab}2L_{ab}$ respectively in the 4-simplex dual to $v$ as $\cT_f^v=\kappa_{ab}\cT_{ab}^v$ and $\widetilde{\cT}^v_f=\kappa_{ab}\widetilde{\cT}_{ab}^v$ respectively. $T_f$ (\resp $\widetilde{T}_f$) is linearly related to $\{\cT^v_f\}_v$ (\resp $\{\widetilde{\cT}^v_f\}_v$) as follows (see the argument after \eqref{eq:tau_chi} and a special example in \cite{Han:2023hbe}).
\be
\exists\, r_f,s_f\in \R \,\text{ \st }\,  T_f= \sum_{v\in f} \cT_f^v +r_f \cdot 2L_f + i\pi s_f \quad \text{and } 
\widetilde{T}_f= \sum_{v\in f} \widetilde{\cT}_f^v -r_f \cdot 2L_f - i\pi s_f
\,,
\label{eq:Tf_and_cT_f}
\ee
where $v\in f$ denotes that the triangle dual to $f$ is shared by the 4-simplices dual to $v$'s. 

\medskip

Let us first consider the derivative of the action \eqref{eq:effective_action} \wrt $2L_f$ for some internal spinfoam face $f=(ab)$, which gives\footnote{{We also have the symmetry $\cT_{ab}=-\cT_{ba}$ as the FN lengths. When one expresses $\cT^v_f$ and $\widetilde{\cT}^v_f$ in \eqref{eq:Seff_d2Lf} explicitly as $\{\cT^v_{ab},\widetilde{\cT}^v_{ab}\}_v$ with $a<b$, a minus sign appears when two annuli of opposite orientations are glued.}}
\be
\f{\partial S}{\partial (2L_f)}
=-\f{i}{2\pi (b^2+1)}\sum_{v\in f} \lb \cT_{f}^v-b^2\widetilde{\cT}_f^v\rb + \f{i}{2\pi}\cF'_f(2L_f)-u_f\,,
\label{eq:Seff_d2Lf}
\ee
where 
$\cF'_f$ is the derivative of $\cF_f$ \wrt $2L_f$. 
Recalling that  $\cF_f(2L_f)$ is a real function quadratic in $2L_f$ as in \eqref{eq:cF}, then
\be
\cF'_f(2L_f)= 2a_f\cdot 2L_f + i\pi b_f\,.
\ee
Using the relations \eqref{eq:Tf_and_cT_f} and parametrizing $T_f$ and $\widetilde{T}_f$ as
\be
T_f=\f{2\pi i}{k}\lb -ib \nu_f-n_f \rb\,, \quad
\widetilde{T}_f=\f{2\pi i}{k}\lb -ib^{-1} \nu_f+n_f \rb\,,\quad
\nu_f\in\bC\,,\quad n_f\in\Z/k\Z\,,
\ee
the critical equation from \eqref{eq:Seff_d2Lf} takes the form
\be
-\f{n_f}{k} + \f{i (r_f+2a_f)}{2\pi} \cdot 2L_f -\f{s_f+b_f}{2} -u_f=0\,.
\ee
Restricting the range of $n_f\in [0,k)$ and fixing the coefficients $(a_f,b_f)$ for each face amplitude, there is only one solution to $n_f$ and $u_f$ as $u_f\in \Z$ while $n_f/k\in [0,1)$. Specially, choosing $a_f=-\f12 r_f$ and $b_f=-s_f$, the solution to $n_f$ and $u_f$ is simply
\be
n_f=0\,,\quad u_f=0\,.
\label{eq:critical_sol_n_f}
\ee

We next consider the geometrical interpretation of the critical solution \eqref{eq:critical_sol_n_f}. 
Given a unique solution to $n_f$, we have
\be
n_f=\f{ik}{2\pi(b^2+1)}\lb T_f-b^2\widetilde{T}_f \rb 
=\f{\nu k\gamma}{4\pi }\varepsilon^{(s)}_f -k N_f\,,
\label{eq:n_f2deficit}
\ee
leading to a unique solution to the dressed deficit angle:
\be
\varepsilon^{(s)}_f=\f{4\pi \nu }{k\gamma}\lb n_f+kN_f \rb \,.
\ee
When we choose the definition of the face amplitude such that the solution \eqref{eq:critical_sol_n_f} to $n_f$ is obtained, we get a constraint similar to the  EPRL model \cite{Bonzom:2009hw,Han:2013hna}:
\be
\varepsilon_f^{(s)} =4\pi N_f/\gamma\,,\quad  N_f\in\Z\,.
\ee
The dressed deficit angle can only take discrete values at the critical point. $N_f$ is fixed by $\varepsilon_f^{(s)}$ from the geometry described by the critical point. In particular 
\be
\varepsilon_f=0
\ee
with $\mathrm{sgn}(V_4^v)=1$ (or $-1$) uniformly satisfy the constraint thus is a critical solution. When $\sgn(V_4^v)=1$ for all $v\in f$, $\varepsilon_f=\sum_{v\in f}\Theta^v_f$ measures the geometrical deficit angle hinged by the triangle dual to $f$. The solution with vanishing $\varepsilon_f$ corresponds to a smooth dS or AdS spacetime since all 4-simplices are constantly curved with consistent constant curvature.

\medskip

Let us summarize the geometrical interpretation of the (real) critical points of the spinfoam amplitude \eqref{eq:spinfoam_amplitude}. When each vertex amplitude describes a nondegenerate constantly curved 4-simplex, these 4-simplices are further glued through boundary tetrahedra pairwise by identifying all the triangle areas and the tetrahedron shapes. Finally, the gluing of 4-simplices gives a vanishing deficit angle hinged by each internal triangle, when the 4-simplices have uniform orientation $\sgn(V_4)$. This means the 4-simplices can be seen as sub-simplices of (the triangulation ${\bf T}(\cM_4)$ of) a constantly curved 4-manifold $\cM_4$. 
Namely, the spinfoam amplitude describes a dS spacetime (when $\Lambda>0$) or an AdS spacetime (when $\Lambda<0$) in the semiclassical regime. We call this the ``(A)dSness'' property of this spinfoam model.

\section{Away from the (A)dS-ness}
\label{sec:hormander}

In Sections \ref{sec:stationary_vertex} and \ref{sec:critical_deficit}, we have only considered the real critical points of the spinfoam amplitude. We can, nevertheless, consider the complex critical points by extending $\mu,\nu,m,n$ in the parametrizations of the phase space coordinates to complex variables\footnote{For clarity, the imaginary parts $\im(\mu)=\alpha$ and $\im(\nu)=\beta$ in previous sections are fixed and are there only for convergent contour integrations so $\mu=\re(\mu)$ and $\nu=\re(\nu)$ are still considered real variables in the amplitude. Here, we extend these real parameters to complex variables. }. 
A complex critical point can be seen as the shift of a real critical point from the real axes to the complex (hyper-)plane. This is given by Hormander's theorem (Theorem 7.7.12 of \cite{hormander2015analysis}, see also Theorem 2.3 of \cite{melin2006fourier}), which we formulate into Theorem \ref{theorem:hormander} below as a special case. 

We first express the spinfoam amplitude $\cZ_{\vec{\hrho}_\partial}(\vec{\alpha}|\vec{j}_b)$ \eqref{eq:spinfoam_amplitude} for a 4-manifold $\cM_4$ in the large-$k$ regime: 
\begin{eqnarray}
\cZ_{\vec{\hrho}_\partial}(\vec{\alpha}|\vec{j}_b)&&
\xrightarrow{k\rightarrow\infty}\cN_{\tot}\sum_{\vec{p}\in\Z^{15V}}\sum_{\vec{u}\in\Z^{5V}}\sum_{\vec{u}_f\in\Z^F}
\left[\prod_{e=1}^{E_{\text{in}}}\int_{\overline{\cM}_{\vec{j}^v_a}}\rd \hrho^{v\in e}_a
\int_{\overline{\cM}_{\vec{j}^{v'}_b}}\rd \hrho^{v'\in e}_b\right] 
\left[\int_{\cC^{\times F_{\In}}_{L_f}} \bigwedge_{f=1}^{F_{\In}}\rd \lb i 2L_f \rb \lb i 2L_f\rb^\fp\right]\cI\,,\nonumber\\
&&\text{with }\,
\cI=
\int_{\cC^{\times40V}_{M_v}}\prod_{v=1}^V\rd M_{v} \, \exp\lb k S\rb\,,
\label{eq:Z_large_k}
\end{eqnarray}
where $\cN_\tot=\lb\f{k}{2\pi} \rb^{F_\In}\lb \f{k}{4\pi^2}\rb^{2E_{\In}}\cN^V$, and the action $S$ is defined in \eqref{eq:effective_action}. 
For the integral $\cI$, where the measure $\rd M_v$ is defined in \eqref{eq:dM}, all the coherent state labels $\{\hrho^v_a\}$ and the FN lengths $\{2L_f\}$ on the internal spinfoam faces are regarded as boundary data, which we collect in a vector $\vec{r}\in\R^m$ of real variables with length $m=10V+F_{\In}$.  $S$ is a function of $n=40V$ real variables $\vec{x}= \{\{\nu_I^v,n_I^v\}_{I=1}^{15},\{\mu_a^v,m_a^v\}_{a=1}^5\}_{v=1}^V\in\R^{n}$, which are the parametrization of $\{\{\fP_I^v,\tfP_I^v\}_{I=1}^{15},\{\cX_a^v,\widetilde{\cX}^v_a\}_{a=1}^5\}_{v=1}^V$. Due to the use of Poisson resummation, $\{n_I^v,m_a^v\}$ are all continuous variables with integration range $[-\delta,k-\delta]$. 
The integral $\cI$ can be approximated in terms of complex critical points as follows.  

\begin{theorem}
\label{theorem:hormander}
Let $\vec{x}_0\in\R^n$ be a real critical point of the action $S(\vec{x},\vec{r})$ defined in \eqref{eq:effective_action} with $\vec{x}$ and $\vec{r}$ defined above where the Hessian is non-degenerate at the critical points, \ie $\left.\det\lb \partial^2_{\vec{x}\vec{x}} S \rb\right|_{\vec{x}=\vec{x}_0}\neq 0$ , then
\be
\re\lb S(\vec{x},\vec{r})\rb\leq 0\,,\quad
\re\lb S(\vec{x}_0(\vec{r}),\vec{r})\rb=0\,,\quad 
\left.\f{\partial S(\vec{x},\vec{r})}{\partial \vec{x}}\right|_{\vec{x}=\vec{x}_0(\vec{r})}=0 \,.
\label{eq:action_property}
\ee
Analytic continue $\vec{x}$ to $\vec{z}=\vec{x}+i\vec{y}\in\bC^n$ near the critical point $\vec{x}_0$ with $|\vec{y}|$ small, and solve $\f{\partial S(\vec{z},\vec{r})}{\partial \vec{z}} =0$ for a complex critical point $\vec{z}_0(\vec{r})$. Then at the critical point $\vec{z}_0(\vec{r})$, there exists some $0<C<\infty$ such that
\be
\re\lb S(\vec{z}_0(\vec{r}), \vec{r}) \rb \leq -C |\im\lb \vec{z}_0(\vec{r})\rb|\,.
\label{eq:decay}
\ee
Suppose that $S(\vec{x},\vec{r})$ has finitely many real critical points $\{\vec{x}_0^{(\alpha)}\}_{\alpha}$, and $\{\vec{z}^{(\beta)}_0\}_\beta$ is a collection of the complex critical points at their neighbourhood ($\beta$ is not necessarily equal to $\alpha$). Then the integral $\cI$ defined in \eqref{eq:Z_large_k} can be approximated as 
\be
\cI = \lb \f{1}{k}\rb^{\f{n}{2}} \sum_{\beta}\f{e^{k S(\vec{z}_0^{(\beta)}, \vec{r})}}{\sqrt{\det\lb -H_{\vec{z}_0^{(\beta)}}/(2\pi)\rb}} \lb1+ O(1/k) \rb\,,\quad
\text{with }\,H_{\vec{z}_0}=\left.\f{\partial^2S(\vec{z},\vec{r})}{\partial{\vec{z}}^2}\right|_{\vec{z}=\vec{z}_0^{(\beta)}}\,.
\label{eq:stationary_phase}
\ee
\end{theorem}
The proof follows \cite{hormander2015analysis}. For self-consistency, we provide proof in Appendix \ref{app:proof}. 
As analyzed in the previous section, a real critical point corresponds to a zero deficit angle hinged by an internal triangle (when $\sgn(V_4)=1$ for all 4-simplices). In contrast, a complex critical point gives a non-zero deficit angle.
At the semi-classical regime, therefore, this theorem states that a real critical point corresponds to an (A)dS geometry while a complex critical point corresponds to a non-(A)dS geometry. 

\begin{figure}[h!]
\centering
\includegraphics[width=0.3\textwidth]{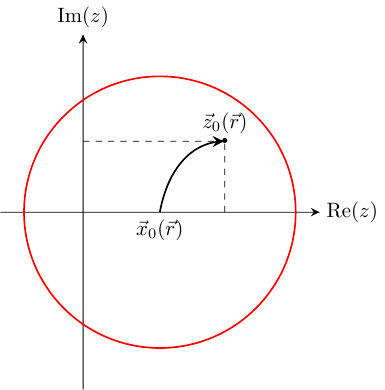}
\caption{A complex critical point $\vec{z}_0(\vec{r})$ in the neighbourhood of a real critical point $\vec{x}_0(\vec{r})$. }
\label{fig:complex_critical_point}
\end{figure}

The complex critical point $\vec{z}_0(\vec{r})$ is an analytic function of the boundary parameter $\vec{r}$ with a real-vector value $\vec{z}_0(\vec{r}_0)=\vec{x}_0$ at $\vec{r}=\vec{r}_0$. $\vec{z}_0(\vec{r})$ deviates from the real space $\R^n$ to $\bC^n$ when $\vec{r}$ deviates from $\vec{r}_0$ with a finite distance, as illustrated in fig.\ref{fig:complex_critical_point}. On the other hand, when the critical point is real, the critical action contributes to an oscillatory phase. 
Eq.\eqref{eq:decay} means that the amplitude decays when the critical point is complex, and that the further the complex critical point is away from the real space, the faster the amplitude decays. At large $k$, the contribution from complex critical points is dominated by the one closest to the real space as others exponentially decay much faster. 

\section{An example: spinfoam amplitude of the $\Delta_3$ 4-complex}
\label{sec:delta3}

The simplest example where one can apply the formalism \eqref{eq:Z_large_k} and the above theorem is the $\Delta_3$ 4-complex, where there is only one internal triangle and it is shared by three 4-simplices, denoted as $v,\,v'$ and $v''$. We will denote elements on $v'$ with primes and those on $v''$ with double primes accordingly in this section. 

\begin{figure}[h!]
\centering
\begin{minipage}{0.45\textwidth}
\centering
\includegraphics[width=\textwidth]{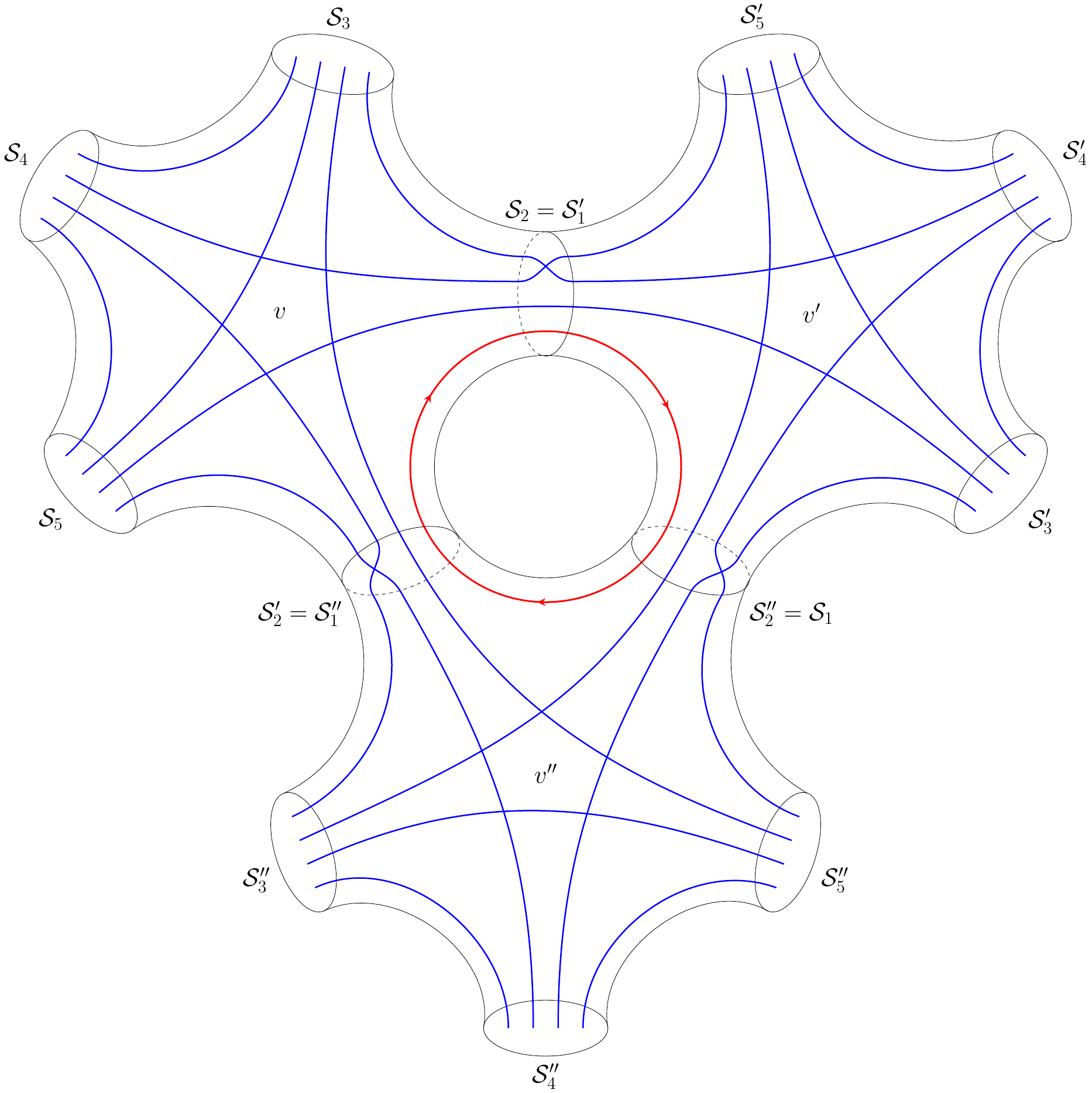}
\subcaption{}
\label{fig:delta3}
\end{minipage}
\qquad
\begin{minipage}{0.4\textwidth}
\centering
\includegraphics[width=\textwidth]{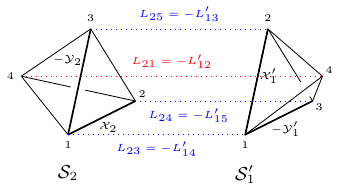}\\[0.3cm]
\includegraphics[width=\textwidth]{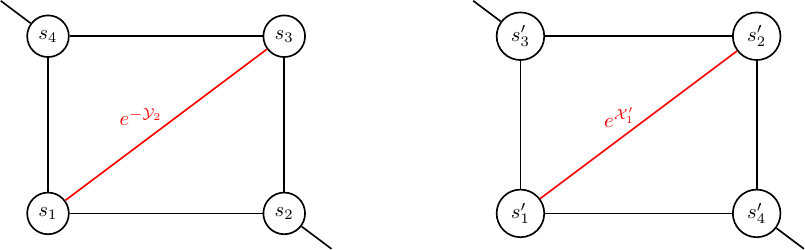}
\subcaption{}
\label{fig:glue}
\end{minipage}
\caption{{\it (a)} Diagram of the 3-manifold corresponding to the $\Delta_3$ 4-complex. The ambient 3-manifold ({\it in black}) has one non-contractable cycle. The (non-intersecting) blue lines denote the annuli and the red loop denotes the torus corresponding to the internal triangle shared by three 4-simplices. The 3-manifold on which the CS amplitude is defined is the graph (composed of the blue lines and the red loop) complement of the ambient 3-manifold. ({\it b}) The upper panel illustrates the gluing of $\cS_2$ and $\cS_1'$.  Numbers 1,2,3,4 label the holes and the dotted lines denote the annuli dressed with FN coordinates that identify the holes pairwise. Edges in thick are dressed with the FG coordinates on the 4-holed spheres. The lower panel illustrates the quadrilateral to define $e^{-\cY_2}$ and $e^{\cX'_1}$ through \eqref{eq:def_Y2_Xp1}. 
}
\label{fig:delta3_glue}
\end{figure}

The diagram of the 3-manifold corresponding to $\Delta_3$ 4-complex is illustrated in fig.\ref{fig:delta3}, which has a similar pattern as the cable diagram of the $\Delta_3$ 4-complex (see \eg \cite{Han:2021kll}).
The patterns (the way the annuli connect different 4-holed spheres) in different $\SG$'s are identical. Denote the gluing, or identifying, of 4-holed spheres by $\sim$, then
the internal triangle comes from 
\be
v\ni\cS_2 \sim \cS_1'\in v'\,,\quad
v'\ni\cS_2'\sim\cS''_1\in v''\,,\quad
v''\ni\cS''_2\sim\cS_1\in v\,.
\label{eq:gluing-spheres}
\ee
 Fig.\ref{fig:glue} illustrates the gluing of $\cS_2$ and $\cS'_1$ as an example. It identifies the FN and FG coordinates from the two simplices as follows (we take $\kappa_{ab}=1$ for $a<b$ and $\kappa_{ab}=-1$ otherwise in this section). 
\be
{e^{L_{21}}=e^{-L'_{12}}\,,\quad
e^{L_{23}}=e^{-L'_{14}}\,,\quad
e^{L_{24}}=e^{-L'_{15}}\,,\quad
e^{L_{25}}=e^{-L'_{13}}\,,\quad
e^{\cX_2}=e^{\cY'_1}\,,\quad
e^{\cY_2}=e^{\cX'_1}\,.}
\label{eq:gluing_constraints}
\ee
Similar relations are true for the other two gluings $\cS_2'\sim \cS''_1$ and $\cS''_2\sim \cS_1$. These constraints result from identifying the framing flags on the glued holes. For instance, as illustrated in fig.\ref{fig:glue}, the exponential FG coordinates $e^{-\cY_2}$ on $\cS_2$ and $e^{\cX'_1}$ on $\cS'_1$ are defined in terms of the framing flags as (see Appendix \ref{sec:app_geo}, especially \eqref{eq:FG_from_flag} and fig.\ref{fig:flag})
\be
e^{-\cY_2}=\f{\la s_4 \w s_1 \ra\la s_3 \w s_2\ra}{\la s_4\w s_3 \ra\la s_1\w s_2\ra}\,,\quad
{e^{\cX'_1}=\f{\la s'_3 \w s'_1 \ra\la s'_2 \w s'_4\ra}{\la s'_3\w s'_2 \ra\la s'_1\w s'_4\ra}\,,}
\label{eq:def_Y2_Xp1}
\ee
where $s_i$ is the framing flag on hole $i$ of $\cS_2$ parallel transported to a common point in $\cS_2$ and $s'_j$ is the framing flag on hole $j$ of $\cS'_1$ parallel transported to a common point in $\cS'_1$. The identification of framing flags $s_1\sim s'_1,\, s_2\sim s'_3,\,s_3\sim s'_2,\,s_4\sim s'_4$ leads to the constraint $e^{\cY_2-\cX'_1}=1$ hence $\cY_2-\cX'_1=0$ with a chosen lift. Other constraints in \eqref{eq:gluing_constraints} can be obtained in the same manner. We collect them in Appendix \ref{app:glue}.

The edge amplitudes for such a gluing is
\be
\cA_e(\hrho_2^v,\hrho_1^{v'}|j_{12},j_{23},j_{24},j_{25},j'_{12},j'_{13},j'_{14},j'_{15})
=\delta_{j_{12},j'_{12}}\delta_{j_{23},j'_{14}}\delta_{j_{24},j'_{15}}\delta_{j_{25},j'_{13}}
\exp[S_e(\hrho_2^v,\hrho^{v'}_1)]\,,
\ee
where $\hrho_2^v=(\zh_2,\hx_2,\yh_2)$ and $\hrho_1^{v'}=(\zh'_1,\hx'_1,\yh'_1)$. The action $S_e(\hrho_2^v,\hrho_1^{v'})$ reads
\begin{eqnarray}
S_e(\hrho_2^v,\hrho_1^{v'})&=&-\f{1}{4\pi}\lb\lb\re(\zh_2)+\im(\zh'_1)\rb^2+\lb\re(\zh'_1)+\im(\zh_2)\rb^2+\lb \xh_2+\yh'_1\rb^2+\lb\yh_2+\xh'_1\rb^2\rb \nonumber  \\
&&+ \f{i}{4\pi}\lb 4\im(\zh'_1)\im(\zh_2) + \xh_2\yh_2+\xh'_1\yh'_1+2\yh_2\yh'_1\rb\,.
\end{eqnarray}
The other two edge amplitudes take the same form except for changing the same elements in $v$ to $v'$ (\resp the same elements in $v$ to $v''$) and those in $v'$ to $v''$ (\resp those in $v'$ to $v$).
At the large-$k$ approximation of the full amplitude for the $\Delta_3$ 4-complex, the following constraints on the FG coordinates are obtained by solving the equations of motion as in Section \ref{sec:edge_amplitude}.
\be
\cX_2-\cY'_1=\cX'_2-\cY''_1=\cX''_2-\cY_1 = 0=\cY_2-\cX'_1=\cY'_2-\cX''_1=\cY''_2-\cX_1\,.
\label{eq:constraints}
\ee 
We can embed the phase spaces for different $\SG$'s into the full phase space for the 3-manifold after gluing. 
Under the standard Poisson bracket, the constraints \eqref{eq:constraints} can be checked to be all first-class.

We are in particular interested in the equations of motion from the variation of the internal FN length $2L_{12}$. It reads
\be
\f{\partial S}{\partial (2L_{12})}
= -\f{i}{2\pi(b^2+1)}\left[ \lb \cT_{12} -b^2\widetilde{\cT}_{12} \rb + \lb \cT'_{12} -b^2\widetilde{\cT}'_{12}\rb +\lb\cT'_{12} -b^2\widetilde{\cT}'_{12} \rb\right] 
+\f{1}{2\pi} \lb 2a_{12}\cdot 2L_{12}+i\pi b_{12}\rb-u_{12}=0\,,
\label{eq:eom_L12}
\ee
where $u_{12}\in\Z$ comes from the Poisson resummation of $j_{12}$ and $a_{12},b_{12}\in \R$ are the coefficients of the face amplitude $\cF_f(2L_{12})=a_{12}\lb 2L_{12}\rb^2+ i\pi  b_{12}\cdot 2L_{12} + c_{12}$. 
The FN twist $T_{12}$ and its tilde sector $\widetilde{T}_{12}$ along the B-cycle of the torus corresponding to this internal triangle are the linear combinations of $2L_{12}$ and its conjugate momenta on three different 4-simplices. Explicitly\footnote{{When gluing the annuli to form the internal torus corresponding to the FN length $L_{12}$, the orientations of the annuli are congruent, as can be seen in fig.\ref{fig:delta3}, hence there is no sign difference for $\cT_{12},\cT'_{12}$ and $\cT''_{12}$ in \eqref{eq:T12}.}},
\be
T_{12}= \cT_{12}+\cT'_{12}+\cT''_{12} + r_{12}\cdot 2L_{12} + i\pi s_{12}\,,\quad
\widetilde{T}_{12}= \widetilde{\cT}_{12}+\widetilde{\cT}'_{12}+\widetilde{\cT}''_{12} - r_{12}\cdot 2L_{12} - i\pi s_{12}\,,\quad 
r_{12},s_{12}\in \R\,.
\label{eq:T12}
\ee
Parametrize the FN twist as
\be
T_{12}=\f{2\pi i}{k}\lb -ib \nu_{12}-n_{12} \rb\,, \quad
\widetilde{T}_{12}=\f{2\pi i}{k}\lb -ib^{-2} \nu_{12}+n_{12} \rb\,,\quad
\nu_{12}\in\R\,,\quad n_{12}\in\Z/k\Z\,.
\label{eq:param_T12}
\ee 
$T_{12}$ and $\widetilde{T}_{12}$ are related to the dressed deficit angle $\varepsilon_{12}^{(s)}$ hinged by the internal triangle through
\be
T_{12}=-\f12\nu \varepsilon_{12}^{(s)} +2\pi i N_{12}\,,\quad
\widetilde{T}_{12}=-\f12\nu\varepsilon_{12}^{(s)} -2\pi i N_{12}\,,\quad N_{12}\in \Z\,.
\label{eq:T12_deficit}
\ee
Define the face amplitude $\cF_f(2L_{12})$ by fixing $a_{12}=-\f12 r_{12}$ and $b_{12}=-s_{12}$, then \eqref{eq:eom_L12} can be simplified to be 
\be
-\f{n_{12}}{k} -u_{12}=0\,,
\ee
whose only solution, when $n_{12}$ is restricted to $[0,k)$, is
\be
n_{12}=0\,,\quad u_{12}=0\,. 
\label{eq:critical_n12}
\ee
This is the real critical solution to the action of (the large-$k$ approximation of) the amplitude for the $\Delta_3$ 4-complex. 
Equating \eqref{eq:param_T12}  with \eqref{eq:T12_deficit}, one gets (recalling that $Q=(b+b^{-1})=2\re(b)$ and $\im(b)=-\gamma\re(b)=-\f{\gamma Q}{2}$)
\be
\f{2\pi b}{k}\nu_{12} \equiv \f{\pi Q}{k}\nu_{12} -i \f{\pi \gamma Q }{k}\nu_{12} =-\f12 \nu \varepsilon_{12}^{(s)} +2\pi iN_{12} 
\quad \Longrightarrow \quad  
\gamma\varepsilon_{12}^{(s)} =4\pi \nu N_{12}\in 4\pi \Z\,.
\ee

Extending the variables $\{\{\nu_I,n_I,\nu'_I,n'_I,\nu''_I,n''_I\}_{I=1}^{15},\{\mu_a,m_a,\mu'_a,m'_a,\mu''_a,m''_a\}_{a=1}^5\}\in \R^{120}$ (at large-$k$ regime) to $\bC^{120}$, the critical solution becomes complex by the Hormander's theorem \ref{theorem:hormander}. The critical solution renders $n_{12}\neq 0$, leading to a non-vanishing deficit angle. Its contribution to the full amplitude is small compared to the real critical solution \eqref{eq:critical_n12} by Theorem \ref{theorem:hormander}.

\section{Conclusion and outlook}
\label{sec:conclusion}

In this paper, we have, in a systematical way, given the complete spinfoam amplitude, composed by vertex amplitudes, edge amplitudes and face amplitudes, for a general 4-complex as the triangulation of a spacetime manifold when a non-vanishing cosmological constant is present. 
It is formulated as finite sums and convergence integrals on the symplectic coordinates of moduli space of $\SL(2,\bC)$ flat connection on copies of $\SG$'s and coherent state labels. 
We have analyzed the critical solutions to the equations of motion at the large-$k$ regime of the full amplitude. The real critical solutions give $\SU(2)$ flat connection on the graph complement of the 3-manifold after gluing different $\SG$'s through boundary 4-holed spheres. 
Each such flat connection determines the geometry of all the 4-simplices as the sub-cells of the 4-complex under study, hence determining the geometry of the full 4-complex. This means that, when the 4-volume of all 4-simplices are positive, the amplitude of this spinfoam model peaks at an (A)dS spacetime depending on the sign of the cosmological constant. 

We have particularly focused on the critical solutions from varying the internal spins, each corresponding to an internal triangle shared by tetrahedra from different 4-simplices, and we observe a similar result as in the EPRL model as follows. 
With the specific definition of the face amplitude, which may vary for different spinfoam faces, and at a specific lift of the phase space coordinate, the real critical point gives a vanishing deficit angle $\varepsilon_f=0$ hinged by each internal triangle,
and different lifts relate to different deficit angles separated by $4\pi/\gamma$. This separation matches that of the EPRL model.  

We have also observed a technical advantage of studying this spinfoam model compared to the EPRL model: the semi-classical approximation formula of the amplitude is simpler in that the infinite many summations coming from each Poisson resummation of internal spin is reduced to a single sum at the large-$k$ regime. 
Apart from that, another advantage of this spinfoam model is the finiteness of amplitude for a general 4-complex, which means no further regularization is needed. With these distinctive features, this spinfoam model extends an invitation for deeper exploration and investigation. We list some of the possible directions to look into below. 
\begin{itemize}
	\item In this work, the full amplitude is constructed by grouping vertex amplitudes, edge amplitudes and face amplitudes by the local amplitude ansatz. Another way to construct the full amplitude is to first write down the CS partition function for the final 3-manifold that corresponds to the 4-complex under study, then couple it with coherent states on the boundary to impose the second-class simplicity constraints. The interpretation of flat connection at the critical points of the action would be better explained if the latter approach is used. However, the difficulty lies in that a symplectic transformation from the FG coordinates on ideal octahedra to suitable coordinates on the final 3-manifold might not exist for a complex 3-manifold. If it exists, it remains the question of whether there is a systematic way to perform such a symplectic transformation for a general 3-manifold.
	\item The complex critical deficit angle is only argued to exist in this paper. 
	Having the complete and concrete spinfoam model, it is interesting the study the complex critical points numerically as is done in the EPRL model \cite{Han:2021kll,Han:2023cen}, and investigate how these complex critical points contribute to the final amplitude. We expect that the finiteness of amplitude would bring benefit to the numerical study. Furthermore, when it involves solving critical solutions to the action, the feature that only polynomial equations are involved (see discussion in \cite{Han:2023hbe} for more details) could also boost the numerical operation compared to that of the EPRL model. 
	\item The form of the face amplitude \eqref{eq:face} is based on the conjecture that the boundary Hilbert space is spanned by some $\fq$-deformed spin network states with $\fq$ a root-of-unity. To investigate if it is true, one needs to construct explicitly the coherent intertwiners spanning such Hilbert space and clarify if there is a canonical bijection between the coherent intertwiners and the boundary data in the spinfoam model. A first step to construct the coherent intertwiners on a homogeneously curved tetrahedron has been initiated in \cite{Han:2024abc}, and these coherent intertwiners span the intertwiner Hilbert space on a curved tetrahedron defined in \cite{Han:2023wiu}. 
	\item An important question is how this spinfoam model relates to the Hamiltonian constraint in the canonical approach. It would be a difficult task for the general setting. 
	To begin with, one can study the truncated model. As the dS spacetime is at the critical points of the spinfoam model, it is interesting to apply it to the cosmological setting by imposing (discretized version of) isotropic and homogeneous conditions. The numerical method could be also helpful for the analysis. 
\end{itemize}

\begin{acknowledgements}
This work receives support from the National Science Foundation through grants PHY-2207763, PHY-2110234, the Blaumann Foundation, the Jumpstart Postdoctoral Program at FAU, and the College of Science Research Fellowship at FAU. \end{acknowledgements}

\appendix
\renewcommand\thesection{\Alph{section}}

\section{Fock-Goncharov coordinates and the Fenchel-Nielsen coordinates}
\label{app:FG_FN}

The FG coordinates $\{\chi^{(a)}_{ij}\}$ that dress the edges in the ideal triangulation of 4-holed spheres on $\SG$ are related to the coordinates $\{\{L_{ab}\}_{a<b},\{\cX_a,\cY_a\}_a\}$ as follows. $\chi_{ij}^{(a)}$ is associated to the edge of the ideal triangulation of $\cS_a$ that shared by $\Oct(i)$ and $\Oct(j)$. 
\be\ba{lll}
\chi^{(1)}_{23}=-\cY_1,&
\chi^{(1)}_{24}=-L_{12}+L_{13}+L_{14}+L_{15}-\cX_{1}+\cY_{1},&
\chi^{(1)}_{25}=\cX_{1},\\[0.15cm]
\chi^{(1)}_{34}=L_{12}-L_{13}-L_{14}+L_{15}+\cX_{1},&
\chi^{(1)}_{45}=L_{12}+L_{13}-L_{14}-L_{15}-\cY_1,&
\chi^{(1)}_{35}=2 L_{14}-\cX_{1}-\cY_1,\\[0.15cm]
\chi^{(2)}_{13}=L_{12}+L_{23}+L_{24}+L_{25}-\cX_{2}+\cY_{2},&
\chi^{(2)}_{14}=-\cY_{2},&
\chi^{(2)}_{15}=\cX_{2},\\[0.15cm]
\chi^{(2)}_{34}=-L_{12}-L_{23}-L_{24}+L_{25}+\cX_{2},&
\chi^{(2)}_{35}=-L_{12}-L_{23}+L_{24}-L_{25}-\cY_{2},&
\chi^{(2)}_{45}=2 L_{23}-\cX_{2}+\cY_{2},\\[0.15cm]
\chi^{(3)}_{12}=L_{13}+L_{23}+L_{34}+L_{35}-\cY_{3},&
\chi^{(3)}_{14}=-2 L_{23}-\cX_{3}+\cY_{3},&
\chi^{(3)}_{15}=\cX_{3},\\[0.15cm]
\chi^{(3)}_{24}=-L_{13}+L_{23}-L_{34}+L_{35}+\cX_{3},&
\chi^{(3)}_{25}=-L_{13}-L_{23}+L_{34}-L_{35}-\cX_{3}+\cY_{3},&
\chi^{(3)}_{45}=-\cY_{3},\\[0.15cm]
\chi^{(4)}_{12}=L_{14}+L_{24}-L_{34}+L_{45}+\cY_{4},&
\chi^{(4)}_{13}=-2 L_{24}-\cX_{4}-\cY_{4},&
\chi^{(4)}_{15}=\cX_{4},\\[0.15cm]
\chi^{(4)}_{23}=-L_{14}+L_{24}+L_{34}+L_{45}+\cX_{4},&
\chi^{(4)}_{25}=-L_{14}-L_{24}-L_{34}-L_{45}-\cX_{4}-\cY_{4},&
\chi^{(4)}_{35}=\cY_{4},\\[0.15cm]
\chi^{(5)}_{12}=L_{15}+L_{25}-L_{35}-L_{45}-\cY_{5},&
\chi^{(5)}_{13}=-2 L_{25}-\cX_{5}+\cY_{5},&
\chi^{(5)}_{14}=\cX_{5},\\[0.15cm]
\chi^{(5)}_{23}=-L_{15}+L_{25}+L_{35}-L_{45}+\cX_{5},&
\chi^{(5)}_{24}=-L_{15}-L_{25}-L_{35}+L_{45}-\cX_{5}+\cY_{5},&
\chi^{(5)}_{34}=-\cY_{5}
\ea
\label{eq:FN2FG}
\ee

\section{Geometrical interpretations of the Fenchel-Nielson coordinates}
\label{sec:app_geo}
 
In this appendix, we review the geometrical interpretations of the FN coordinates, namely that an FN length encodes the area of the boundary curved triangles shared by two tetrahedra and that its dual FN twist encodes the dihedral angle between the two tetrahedra hinged by the same triangle. These geometrical interpretations have been derived in detail in \cite{Haggard:2014xoa} and used in \cite{Han:2021tzw} (see also \cite{Haggard:2015nat,Haggard:2015sl,Han:2015gma}). We only sketch the derivation here. 

We start by identifying the geometrical interpretation of the FN lengths. To this end, we first review the definition of the FG coordinates from framed flat $\SL(2,\bC)$ connections on the ideal triangulation of an $n$-holed sphere \cite{Fock:2003alg,Gaiotto:2009hg,Dimofte:2013lba,Han:2015gma,Han:2021tzw}. 
In the ideal triangulation, each hole of the sphere is triangulated to a cusp boundary $D$, where we associate a framing flag field $s$ satisfying
\be
\rd s(\fp)=As(\fp)\,, \quad \forall\,\fp\in D\,,
\ee
where $A$ is an $\SL(2,\bC)$ flat connection. Therefore, $s(\fp)$ is an eigenvector of the $\SL(2,\bC)$ holonomy of $A$ along a loop surrounding $D$ based at point $\fp$. It can be viewed as a $\bC^2$ vector field when the holonomy is expressed in the fundamental representation. 

\begin{figure}[h!]
\centering
\includegraphics[width=0.25\textwidth]{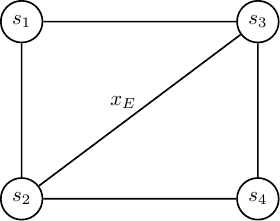}
\caption{A quadrilateral in a 2D ideal triangulation to define FG coordinate $x_E$ in terms of the framing flags $\{s_i\}_{i=1,\cdots,4}$ on four holes (represented by circles) parallel transported to a common point by \eqref{eq:FG_from_flag}. }
\label{fig:flag}
\end{figure}

There are in total $3(n-2)$ edges in the ideal triangulation of an $n$-holed sphere (which does not include the added edges from truncated vertices). Each edge is shared by two triangles and hence can be seen as the diagonal edge of a quadrilateral as shown in fig.\ref{fig:flag}. On each vertex $v_i\, (i=1,\cdots,4)$, or equivalently a cusp boundary $D_i$, of the quadrilateral, there is a framing flag. Parallel transport all four framing flags to a common point within the quadrilateral and label the resulting framing flag transporting from $D_i$ as $s_i$. Referring to the relative locations of the edge $E$ and vertices, the FG coordinate $x_E$ on $E$ is defined as
\be
x_E=\frac{\left\langle s_1 \wedge s_2\right\rangle\left\langle s_3 \wedge s_4\right\rangle}{\left\langle s_1 \wedge s_3\right\rangle\left\langle s_2 \wedge s_4\right\rangle}\,,\quad
\text{with }\, s_i=\lb s_i^0, s_i^1\rb^\top\in\bC^2\,,\quad 
\left\langle s_i \wedge s_j\right\rangle=s_i^0 s_j^1-s_i^1s_j^0\,,
\label{eq:FG_from_flag}
\ee
which is invariant for complex rescaling of any flag $s_i$. 

An $\SL(2,\bC)$ holonomy along a closed loop can be calculated by the so-called {\it snake rule}\footnote{We refer to \cite{Dimofte:2013lba} for a detailed description of the snake rules. See also appendices of \cite{Han:2023hbe}. } followed by a normalization. By the snake rule, one can calculate that the holonomy $O^{(a)}_i$ on $\cS_a$ around hole $i\,(i=1,\cdots,4)$ connected to a hole of $\cS_b$ through the annulus $(ab)$ on $\partial(\SG)$ is conjugated to a diagonal matrix. That is,  
\be
O^{(a)}_i=M \, \diag(e^{L_{ab}},e^{-L_{ab}}) \, M^{-1}\in\SL(2,\bC)\,,\quad O^{(a)}_i,M\in\SL(2,\bC)\,,
\ee
where $e^{L_{ab}}$ is (the square root of) the exponential FN length on $(ab)$. Throughout this appendix, we assume $a<b$. 
$\{O^{(a)}_i\}_{i=1,\cdots,4}$ are by definition the holonomies of flat connections on $\cM_\Flat(\cS_a,\SL(2,\bC))$ since they are calculated by the snake rule. Therefore, they satisfy the closure condition $O^{(a)}_4O^{(a)}_3O^{(a)}_2O^{(a)}_1=\id$ when all the holonomies are based at the same point on $\cS_a$. 
When the first-class simplicity constraints are imposed strongly on the FN coordinates such that 
\be
L_{ab}= -2\pi i \f{j_{ab}}{k}, \quad a<b\,,\quad\text{with } \,j_{ab}=0,\f12,\cdots, \f{k-1}{2}
\label{eq:L-to-j}
\ee
as described in Section \ref{sec:review}, 
each $O^{(a)}_i$ is conjugated to an $\SU(2)$ holonomy, denoted as $H^{(a)}_i$, and $\{H^{(a)}_i\}_{i=1,\cdots,4}$ also satisfy the closure condition $H^{(a)}_4H^{(a)}_3H^{(a)}_2H^{(a)}_1=\id$ 
when they have the same base point, denoted as $\fp_a$, on $\cS_a$. 

According to the curved tetrahedron reconstruction theorem proven in \cite{Haggard:2015ima}, the set of $\SU(2)$ holonomies $\{H^{(a)}_i\}_{i=1,\cdots,4}$ satisfying the closure condition uniquely identifies a homogeneously curved tetrahedron, denoted as $T_a$, such that the area and the normal vector of each of its boundary triangles, denoted as $t_i^{(a)}\,(i=1,\cdots,4)$, can be read from $H^{(a)}_i$ in the following way. Diagonalize $H^{(a)}_i$ with the matrix $M(\xi_i)\in\SU(2)$ constructed with the eigenvectors, also called the spinors, of $H^{(a)}_i$ such that
\be
H^{(a)}_i = M(\xi_i)\, \diag(e^{-2\pi i  \f{j_{ab}}{k}},e^{2\pi i \f{j_{ab}}{k}}) \, M(\xi_i)^{-1}\,,\quad
M(\xi_i)=\lb \xi_i,\,J\xi_i \rb\,,\quad
\text{with }\,\left|\ba{rl}
\xi_i=&\lb \xi_i^0,\xi_i^1 \rb^\top\\[0.1cm]
 J\xi_i=&\lb -\bar{\xi}_i^1,\bar{\xi}_i^0 \rb^\top
 \ea\right..
 \label{eq:diagonalize_H}
\ee
We will also use the notation $|\xi\ra\equiv \xi_i$ and $|\xi_i]:=J\xi_i$ and $\la \xi_i|,[\xi_i|$ represent their transpose conjugates respectively. 
$|\xi_i\ra$ and $|\xi_i]$ are orthonormal in the sense that $[\xi_i|\xi_i\ra=\la\xi_i|\xi_i]=0$ and they are both normalized: $\la\xi_i|\xi_i\ra=[\xi_i|\xi_i]=1$.
The area $\fa_{ab}$ and the normal $\hat{n}_{ab}$ of $t_i^{(a)}$ calculated at $\fp_a$ is
\be
\fa_{ab}=
\begin{cases}
\f{12\pi j_{ab}}{|\Lambda|k}\,,\quad & \text{if } j_{ab}\in [0,\f{k}{4})\\
\f{6\pi}{|\Lambda|}-\f{12\pi j_{ab}}{|\Lambda|k}\,,\quad & \text{if } j_{ab}\in [\f{k}{4},\f{k}{2})
\end{cases}\,,\quad
\hat{n}_{ab}=
\begin{cases}
\la\xi_i| \vec{\sigma}| \xi_i\ra\,,\quad & \text{if } j_{ab}\in [0,\f{k}{4})\\
-\la\xi_i| \vec{\sigma}| \xi_i\ra& \text{if } j_{ab}\in [\f{k}{4},\f{k}{2})
\end{cases}\,,
\label{eq:area_quantum}
\ee
where $\vec{\sigma}=(\sigma^1,\sigma^2,\sigma^3)$ is a vector of Pauli matrices. 
The outward-pointing normal $\hfn_{ab}$ to $t_i^{(a)}$ is different from $\hat{n}_{ab}$ by a sign factor $\nu=\sgn(\Lambda)$, namely
\be
\hfn_{ab}=\nu \hat{n}_{ab}\,.
\ee
This is because the normalized eigenvector $\xi_i$ is the same for holonomies around a spherical triangle (corresponding to $\nu=+$) with eigenvalue, say $\lambda$, and a hyperbolic one (corresponding to $\nu=-$) with eigenvalue $\lambda^{-1}$. For each of the all four triangles in a tetrahedron, either the area is related to $j_{ab}$ in the first or second option in \eqref{eq:area_quantum} is determined by the triple product $(\hat{n}_i\times \hat{n}_j)\cdot \hat{n}_k\stackrel{!}{=}\nu$ for any set of three triangles in a tetrahedron. On the other hand, $(\hfn_i\times \hfn_j)\cdot \hfn_k>0$ for either $\nu$ \cite{Haggard:2015ima}.

$\xi_i$ is in fact the (normalized) framing flag $s_i$ parallel transported to the base point $\fp_a$  \ie $\xi_i=\f{s_i}{||s_i||}$. Therefore, $\{\xi_i\}_{i=1,\cdots,4}$ can be used to define the FG coordinates as in \eqref{eq:FG_from_flag}. 

The FN lengths admit the symmetry $L_{ab}=-L_{ba}$,
which geometrically means that $t_i^{(a)}$ and the triangle $t_j^{(b)}$ on $T_b$ corresponding to some $H_j^{(b)}$, such that hole $i$ of $\cS_a$ and hole $j$ of $\cS_b$ are connected by annulus $(ab)$, share the same area $\fa_{ab}$. We can also diagonalize this $H_j^{(b)}$: 
\be
H^{(b)}_j = M(\xi'_j)\, \diag(e^{2\pi i \f{j_{ab}}{k}},e^{-2\pi i \f{j_{ab}}{k}}) \, M(\xi'_j)^{-1}\,,\quad
M(\xi'_j)=\lb \xi'_j,\,J\xi'_j \rb\,, 
\label{eq:Hb}
\ee
where $|\xi'_j\ra\equiv\xi'_j$ and $|\xi'_j]\equiv J\xi'_j$ are defined in the same way as $\xi_i$ and $J\xi_i$ in \eqref{eq:diagonalize_H}. 
The normal of $t_j^{(b)}$, which is defined as $\hat{n}_{ba}=\la\xi'_j|\vec{\sigma}|\xi'_j\ra=\nu\hfn_{ba}$ if $j_{ab}\in[0,k/4)$ while $\hat{n}_{ba}=-\la\xi'_j|\vec{\sigma}|\xi'_j\ra=\nu\hfn_{ba}$ if $j_{ab}\in[k/4,k/2)$, is in general different from $\hat{n}_{ab}$ since they are calculated in different tetrahedron local frames. 
We can also drop the label for holes and denote $H_{ab}\equiv H^{(a)}_i$ and $H_{ba}\equiv H^{(b)\,-1}_j$, 
 whose parametrizations \eqref{eq:diagonalize_H} and \eqref{eq:Hb} can be equivalently written as \cite{Haggard:2015nat}
\be
H_{ab}= e^{\f{\Lambda}{3}\fa_{ab}\hfn_{ab}\cdot \vec{\tau}}\,,\quad
H_{ba}= e^{-\f{\Lambda}{3}\fa_{ab}\hfn_{ba}\cdot \vec{\tau}}\,,
\ee
where $\vec{\tau}=\f{1}{2i}\vec{\sigma}$. (Note that $e^{L_{ba}}$ is the eigenvalue of $H_j^{(b)}$ instead of the eigenvalue of $H_{ba}$.)
$H_{ab}$ and $H_{ba}$ are related through conjugation by an $\SL(2,\bC)$ element, denoted as $G_{ab}$:
\be
H_{ab}= G_{ab} \, H_{ba} \, G_{ab}^{-1}\,.
\label{eq:Hi2Hj}
\ee
$G_{ab}$ describes the parallel transport of the reference frame of $T_a$ to $T_b$. 
There is no canonical choice for $G_{ab}$ and each describes the parallel transport along a path on the annulus $(ab)$ of the reference frame of $T_a$ to $T_b$\footnote{An apparent example is $G_{ab}=M'(\xi_i)M(J\xi'_j)^{-1}\in\SU(2)$. However, complex rescalings of $\xi_i$ and $\xi'_j$ \st $|\xi_i\ra\rightarrow \lambda  |\xi_i\ra,\,|\xi_i]\rightarrow\lambda^{-1}|\xi_i]$ and $|\xi_j\ra\rightarrow \lambda' |\xi_j\ra,\,|\xi'_j]\rightarrow{\lambda'}^{-1}|\xi'_j]$ with any $\lambda,\lambda'\in\bC\backslash\{0\}$ preserve the relation \eqref{eq:Hi2Hj}. }. 

By the factorizations \eqref{eq:diagonalize_H} and \eqref{eq:Hb} of $H_{ab}$ and $H_{ba}$ respectively, \eqref{eq:Hi2Hj} can be reformulated as
\be
\mat{cc}{\lambda_{ab}&0\\0&\lambda^{-1}_{ab}} M_a^{-1}G_{ab} M_b = M_a^{-1}G_{ab} M_b \mat{cc}{\lambda_{ab}&0\\0&\lambda^{-1}_{ab}}\,,\quad
\lambda_{ab}=e^{-i\f{|\Lambda|}{6}\fa_{ab}}\,,
\ee
where $M_a=M(\xi_i)$ and $M_b=M(\xi'_j)$\,. This means $M_a^{-1}G_{ab} M_b\in \bU(1)$ and can be parametrized as 
\be
M_a^{-1}G_{ab} M_b=\mat{cc}{\gamma'_{ab}&0\\0&\gamma^{'\,-1}_{ab}}\,,\quad
\gamma'_{ab}=\begin{cases}
    \gamma_{ab}\,,\quad & \text{if } j_{ab}\in[0,\f{k}{4})\\
    \gamma^{-1}_{ab}\,,\quad &\text{if } j_{ab}\in[\f{k}{4},\f{k}{2})
\end{cases}\,,\quad
\gamma_{ab}=e^{\psi_{ab}+i\theta_{ab}}\,,\quad
\psi_{ab}\in\R\,,\,\theta_{ab}\in[0,2\pi)\,.
\label{eq:def_tau}
\ee
In the rest of the derivation, we will eliminate the labels of holes on $T_a$ and $T_b$. When hole $i$ of $T_a$ is glued to hole $j$ of $T_b$ through annulus $(ab)$, we denote 
\be
\xi_{ab}:=\begin{cases}
    \xi_i\,,\quad & \text{if } j_{ab}\in[0,\f{k}{4})\\
    J\xi_i \,,\quad &\text{if } j_{ab}\in[\f{k}{4},\f{k}{2})
\end{cases}\,,\quad
\xi_{ba}:=\begin{cases}
   \xi'_j \,,\quad & \text{if } j_{ab}\in[0,\f{k}{4})\\
   J\xi'_j \,,\quad &\text{if } j_{ab}\in[\f{k}{4},\f{k}{2})
\end{cases}\,.
\ee
Then \eqref{eq:def_tau} means that the spinors $\xi_{ba}$ and $\xi_{ab}$ are related by parallel transportation in the manifold of $\SL(2,\bC)$ followed by a rescaling. Explicitly, 
\be
|\xi_{ab}\ra = \gamma^{-1}_{ab} G_{ab}| \xi_{ba}\ra\,,\quad
|\xi_{ab}] = \gamma_{ab} G_{ab} |\xi_{ba}]\,.
\label{eq:xi2xi}
\ee
One of these formula gives the parallel transport from $\xi_i$ to $\xi'_j$, which means $\xi_i$ can be used as the framing flag to define flat connection on the whole boundary $\partial(\SG)$. 

Denote by $\Ht_{ab}\equiv \Ht_{ba}\in\SL(2,\bC)$ for the holonomy along the meridian loop of the annulus $(ab)$ in the fundamental group $\pi_1(\SG)$. The set $\{\Ht_{ab}\}_{a<b}$ of 10 holonomies are the $\SL(2,\bC)$ representations of the generators of $\pi_1(\SG)$. It has been proven in \cite{Haggard:2014xoa} that $\pi_1(\SG)$ is isomorphic to the fundamental group $\pi_1(\text{4-simplex})$ of the 4-simplex bounded by $S^3$. Therefore, $\{\Ht_{ab}\}$ also represent the generators of $\pi_1(\text{4-simplex})$ and describe the $\SL(2,\bC)$ flat connections on the 4-simplex. 
 
\begin{figure}[h!]
\centering
\begin{minipage}{0.3\textwidth}
\centering
\includegraphics[width=0.8\textwidth]{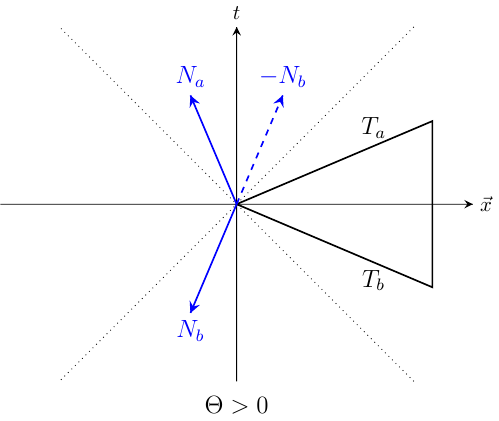}
\subcaption{}
\label{fig:thin_wedge}
\end{minipage}
\begin{minipage}{0.3\textwidth}
\centering
\includegraphics[width=0.8\textwidth]{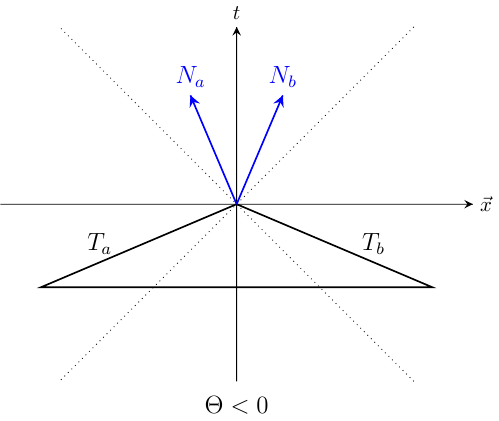}
\subcaption{}
\label{fig:thick_wedge}
\end{minipage}
\caption{Two spacelike tetrahedra $T_a$ and $T_b$ forming a wedge (2 spacial dimensions are reduced). $N_a$ and $N_b$ are the outward-pointing normal to $T_a$ and $T_b$ respectively. {\it (a)} A thin wedge with dihedral angle $\Theta_{ab}>0$. {\it (b)} A thick wedge with dihedral angle $\Theta_{ab}<0$.}
\label{fig:dihedral}
\end{figure}

Parametrize $G_{ab}=g_a^{(b)\,-1}g_b^{(a)}$ such that $g_a^{(b)}$ and $g_b^{(a)}$ (both depending on the tetrahedra $T_a$ and $T_b$, \ie $g_a^{(b)}\neq g_a^{(c)}$ for $b\neq c$)\footnote{We refer to \cite{Haggard:2014xoa} for explicit example for $g_a^{(b)}$ such that $g_a^{(b)}\neq g_a^{(c)}$ when $b\neq c$. } are the gauges relating $\Ht_{ab}$ to $H_{ab}$ and $H_{ba}$ respectively by
\be
\Ht_{ab}=g_a^{(b)}\,H_{ab}\,g_a^{(b)\,-1}=g_b^{(a)}\,H_{ba}\,g_b^{(a)\,-1}\,.
\ee
$g_a$ (\resp $g_b$) then represents changing the local frame of $T_a$ (\resp $T_b$) to a common reference frame of all 5 tetrahedra. 
Equivalently speaking, it corresponds to parallel transporting the base point of the fundamental group generators of $\cS_a$ for all $a=1,\cdots,5$ to a common point on the 3-manifold $\SG$. 
In each tetrahedron local frame, $T_a$ is spacelike hence the 4D normal is $U_a=(1,0,0,0)^\top$. Denote by $\bLambda_{ab}$ for the 4-vector representation of $G_{ab}$, then from \eqref{eq:def_tau}, 
\be
\bLambda_{ab}=\bR_a \, e^{2\psi_{ab}\bK^3-2\theta_{ab} \bJ^3} \, \bR^{-1}_b
\equiv\bR_a \, e^{2\psi_{ab}\bK^3}e^{-2\theta_{ab} \bJ^3} \, \bR^{-1}_b\,,
\label{eq:bLambda}
\ee
where $\bR_a=\bR(\xi_{ab})$ and $\bR_b=\bR(\xi_{ba})$\footnote{$\bR(\xi)=\mat{cc}{1&0\\0&R(\xi)}$ where $R(\xi)$ is a $3\times3$ matrix with elements $R^i_{\phantom{i}j}(\xi)=\f12\tr(\sigma^j M(\xi) \sigma^i M(\xi)^{-1})$.} 
 are the rotation matrices representing $M_a$ and $M_b$ respectively in $4\times4$ matrices, and $\vec{\bK}=(\bK^1,\bK^2,\bK^3)$, $\vec{\bJ}=(\bJ^1,\bJ^2,\bJ^3)$ are the boost and rotation generators of the proper orthochronous Lorentz group $\SO(1,3)^+\cong \PSL(2,\bC)$ written as $4\times4$ matrices. They satisfy the commutation relations
\be
\left[\bJ^i,\bJ^j\right]=\epsilon^{i j}{ }_k \bJ^k\,, \quad
\left[\bK^i, \bK^j\right]=-\epsilon^{i j}{ }_k \bJ^k \,,\quad
\left[\bK^i, \bJ^j\right]=\epsilon^{i j}{ }_k \bK^k\,.
\ee
$\bLambda_{ab}$ measures the hyper-dihedral angle $\Theta_{ab}$ between $T_a$ and $T_b$ through 
\be
-\cosh \Theta_{ab} =\eta_{IJ} u_a^I \bLambda_{ab} u^J_b\,,
\label{eq:dihedral_def}
\ee
where $u_a=u_b=(1,0,0,0)^\top$ are the normals of $T_a$ and $T_b$ respectively in the local reference frame. The existence of the minus sign in \eqref{eq:dihedral_def} is because the dihedral angle is defined to be positive for a thin wedge and negative for a thick wedge. Two spacelike tetrahedra in a 4-simplex form a thin wedge if one of the outward-pointing normals (relative to the 4-simplex) is future-pointing while the other is past-pointing; they form a thick wedge if their outward-pointing normals are both future-pointing or past-pointing (see fig.\ref{fig:dihedral}).
As $\bR_a$ and $e^{-2\theta_{ab} \bJ^3}\bR^{-1}_b$ are rotation matrices, they stablize $u_a$ and $u_b$ respectively. Therefore, \eqref{eq:dihedral_def} can be simplified to be 
\be
\cosh \Theta_{ab} = \cosh 2\psi_{ab}
\quad\Longrightarrow\quad
\Theta_{ab}=\pm 2\psi_{ab}\,.
\ee

It remains to fix the sign of the correspondence, which is done by the following consideration.
Let $N_a$ and $N_b$ be the outward-pointing normals of $T_a$ and $T_b$ respectively in a {\it common} frame, which could be future-pointing or past-pointing. Denote $U_a$ and $U_b$ to the the corresponding future-pointing normals. Then $U_a=\pm N_a$ and $U_b=\pm N_b$.
 The boost from $U_a$ to $U_b$ encodes the hyper-dihedral angle in the transformation matrix $\bL_{ab}\in\SO(1,3)^+$ such that $\bL_{ab}U_a=U_b$. Explicitly\footnote{$\bK_i$ and $\bK^i$ are undistinguished in this paper. Same for $\bJ_i$ and $\bJ^i$.},
\be
\bL_{ab}=e^{ |\Theta_{ab}| \f{U_a\w U_b}{|U_a\w U_b|}}
\equiv e^{|\Theta_{ab}|\f{U_a^{[I}U_b^{J]}\cJ_{IJ}}{|U_a\w U_b|}}\,,\quad
\text{with } \, \cJ_{0i}=\bK_{i}\,,\quad \cJ_{ij}=\epsilon_{ij}^{\phantom{ij}k}\bJ_{k}\,.
\label{eq:Lab}
\ee
Let us check that $\bL_{ab}$ defined as such does transport $U_a$ to $U_b$. With no loss of generality, choose the coordinate system such that $U_a=(1,0,0,0)^\top$ and $N_b$ is on the $tx$-plane. Then $U_b=(\cosh\Theta_{ab},\sinh|\Theta_{ab}|,0,0)^\top$\footnote{The minus sign comes from our convention for the dihedral angle of a thick or thin wedge as illustrated in fig.\ref{fig:dihedral}.}. As $U_a\w (U_b+cU_a)\equiv U_a\w U_b\,,\,\forall\,c\in\R$, we can choose a vector $U_b'$ as the (normalized) linear combination of $U_a$ and $U_b$ and is orthogonal to $U_a$. That is, let $U'_b=(0,1,0,0)^\top$ and hence $|\Theta_{ab}|U_a\w U'_b= |\Theta_{ab}|U_a^0{U_b'}^1\cJ_{01}=|\Theta_{ab}|\bK^1$, leading to 
\be
\bL_{ab}=\mat{cccc}{\cosh\Theta_{ab} & \sinh |\Theta_{ab}| & 0 & 0 \\
\sinh|\Theta_{ab}| & \cosh\Theta_{ab} & 0 & 0 \\
0 & 0 & 1 & 0\\ 0 & 0 & 0 & 1}
\quad \Longrightarrow\quad
\bL_{ab} U_a=U_b\,.
\ee
Note that $|\Theta_{ab}|U_a\w U_b= -\Theta_{ab} N_a\w N_b$ since when $T_a$ and $T_b$ form a thin wedge, $\Theta_{ab}>0$ and the time component of either $N_a$ or $N_b$ is negative, while both time components take the same sign when $T_a$ and $T_b$ form a thick wedge and $\Theta_{ab}<0$ ({\it r.f.} fig.\ref{fig:dihedral}). 

To proceed, we first show identity $\f{N_a\w N_b}{|N_a\w N_b|} =\nu \sgn(V_4)\hat{n}_{ab}\cdot\vec{\bK}$ in the following.
Consider a homogeneously curved spacetime $(\cM_4,g_{\mu\nu})$.
Let $e_I^\mu(\fb)$ be a generic orthonormal frame at a vertex $\fb$ of the triangle $f_{ab}$ shared by $T_a$ and $T_b$, and $\epsilon_{\alpha\beta\mu\nu}$ be an arbitrary volume element on $\cM_4$ compatible with $g_{\mu\nu}\equiv\eta_{\mu\nu}e_I^\mu e_I^\nu$. Then $\sgn(V_4)$ is defined by the compatibility between $\epsilon_{\alpha\beta\mu\nu}$ and $e^I_\alpha$:
\be
\epsilon=\sgn(V_4) e^0\w e^1 \w e^2 \w e^3\,.
\ee
The volume element of $f_{ab}$ is then defined by $\epsilon_{\alpha\beta}=\epsilon_{\alpha\beta\mu\nu}N_a^\mu(\fb)N_b^\nu(\fb)$ with $N^\mu=N^Ie_I^\mu$. Then the following relation holds.
\be
\f{\star \lb N_a(\fb)\w N_b(\fb)\rb}{|\star \lb N_a(\fb)\w N_b(\fb)\rb |}=\sgn(V_4)\lb \epsilon^{\alpha\beta}e_\alpha e_\beta \rb_{ab}(\fb)\,.
\label{eq:triangle_volume}
\ee
In the local frame of $T_a$ whose timelike normal is $u=(1,0,0,0)^\top$, $\epsilon^{\alpha\beta}e_\alpha e_\beta=\hat{\fn}_{ab}\cdot \vec{\bJ}$, which can be viewed as an $\sl(2,\bC)$ element. $\sl(2,\bC)$ can be viewed as a 6D algebra with real generators $\vec{\bJ}=\vec{\tau}$ and $\vec{\bK}=-i\vec{\tau}$. Then the duality map $\star$ acts as $\star\vec{\bJ}=-\vec{\bK}$ and $\star\vec{\bK}=\vec{\bJ}$. Therefore, in the frame of $T_a$, 
\be
\star\lb \epsilon^{\alpha\beta}e_\alpha e_\beta\rb = -\hat{\fn}_{ab}\cdot \vec{\bK} =-\nu\hat{n}_{ab}\cdot\vec{\bK}\,.
\label{eq:triangle_boost}
\ee
Combining \eqref{eq:triangle_volume} and \eqref{eq:triangle_boost}, we conclude that 
\be
\f{N_a\w N_b}{|N_a\w N_b|}=\nu \sgn(V_4)\hat{n}_{ab}\cdot\vec{\bK}\,.
\ee
We then can re-express \eqref{eq:Lab} as
\be
\bL_{ab}(a)
\equiv \exp\lb- \nu\sgn(V_4) \Theta_{ab}\hat{n}_{ab}\cdot\vec{\bK} \rb\,.
\label{eq:bL}
\ee
On the other hand, $\bLambda_{ab}$ \eqref{eq:bLambda} can also be rewritten as
\be
\bLambda_{ab}=\lb \bR_a \, e^{2\psi_{ab}\bK^3}\bR^{-1}_a\rb\lb\bR_a e^{-2\theta_{ab} \bJ^3} \, \bR^{-1}_b\rb 
=\exp \lb 2\psi_{ab}\hat{n}_{ab}\cdot\vec{\bK} \rb \bR'\,,
\label{eq:bLambda_2}
\ee
where we have used the fact that $\bR_a\hat{z}=\hat{n}_{ab}$ 
and that $\bR'=\bR_a e^{-2\theta_{ab} J^3} \, \bR^{-1}_b$ is a pure rotation. 
Both $\bLambda_{ab}$ and $\bL_{ab}$, now written in the frame of $T_b$ can transform the normal $N_b$ to $N_a$, which means their boost parts must agree, \ie
\be
\exp\lb- \nu\sgn(V_4) \Theta_{ab}\hat{n}_{ab}\cdot\vec{\bK} \rb=\exp \lb 2\psi_{ab}\hat{n}_{ab}\cdot\vec{\bK} \rb 
\quad\Longrightarrow\quad
\Theta_{ab}=-2\nu\sgn(V_4) \psi_{ab}\,.
\ee 

Let us finally relate the hyper-dihedral angle to the FN twist. 
The definition of the $\SL(2,\bC)$ FN twist along an annulus $(ab)$ depends on the choice of another two auxiliary holes on $\cS_a$ and another two auxiliary holes on $\cS_b$, or effectively depends on the choice of a path on $(ab)$ connecting $\cS_a$ and $\cS_b$. Let $s_{ab}$ be the framing flag on $(ab)$ and $s_{ac},\,s_{ad}$ (\resp $s_{be},\,s_{bf}$) be the framing flags on the other two holes of $\cS_a$ (\resp $\cS_b$) which connect to $\cS_c$ and $\cS_d$ (\resp $\cS_e$ and $\cS_f$) respectively. Then the (exponential) $\PSL(2,\bC)$ FN twist is defined as
\be
\tau^2_{ab}=- \f{\la s_{be}(\fp_b)\w s_{bf}(\fp_b) \ra}{\la s_{be}(\fp)\w s_{ab}(\fp) \ra\la s_{bf}(\fp)\w s_{ab}(\fp) \ra}
\f{\la s_{ac}(\fp)\w s_{ab}(\fp)\ra\la s_{ad}(\fp)\w s_{ab}(\fp) \ra}{\la s_{ac}(\fp_a)\w s_{ad}(\fp_a)\ra}\,,
\label{eq:FN-twist_def}
\ee
where $\fp_a\in\cS_a$, $\fp_b\in\cS_b$, and $\fp$ is a common point for evaluating $s_{ab}\w s',\,\forall\,s'$ . $\tau^2_{ab}$ is indeed invariant under the rescaling of framing flags. 
As we have observed, the role of framing flags can be played by the spinors when they are defined on a common point on the 4-holed sphere. 
Let us choose $\fp=\fp_b$. In order to evaluate the second ratio in \eqref{eq:FN-twist_def} at $\fp_a$, we need to parallel transport the framing flags with $G_{ab}$: $s(\fp_b)=G_{ab}^{-1}s(\fp_a)$. Then the second ratio in \eqref{eq:FN-twist_def} can be re-expressed as
\be
\f{\la G_{ab}^{-1}s_{ac}(\fp_a)\w s_{ab}(\fp_b)\ra\la G_{ab}^{-1}s_{ad}(\fp_{a})\w s_{ab}(\fp_b) \ra}{\la s_{ac}(\fp_a)\w s_{ad}(\fp_a)\ra}
=\f{\la G_{ab}^{-1} \xi_{ac}\w \xi_{ba}\ra\la G_{ab}^{-1} \xi_{ad}\w \xi_{ba} \ra}{[ \xi_{ac}| \xi_{ad}\ra}=\gamma_{ab}^{2} \f{[\xi_{ac}| \xi_{ab}\ra[ \xi_{ad}| \xi_{ab} \ra}{[ \xi_{ac}|\xi_{ad}\ra}\,,
\label{eq:s2xi}
\ee
where we have used the fact that the produce $\la\, \cdot\,\w\,\cdot\,\ra$ is $\SL(2,\bC)$ invariant hence $\la G^{-1}_{ab} \xi'\w \xi_{ba}\ra=\la  \xi'\w G_{ab}\xi_{ba}\ra=\gamma_{ab}[ \xi'| \xi_{ab}\ra$ for any $\xi'$ by \eqref{eq:xi2xi}. 
We then lift $\tau^2_{ab}$ to an $\SL(2,\bC)$ FN twist by taking its positive square root $\tau_{ab}$. 
We can, therefore, express $\tau_{ab}$ in terms of the spinors as
\be
\tau_{ab}=\gamma_{ab}\sqrt{\chi_{ab}(\xi)}\equiv e^{-\f12 \nu\sgn(V_4)\Theta_{ab}+i\theta_{ab}}\sqrt{\chi_{ab}(\xi)}\,,\quad
\chi_{ab}(\xi)=-\f{[\xi_{be}| \xi_{bf} \ra}{[ \xi_{be}| \xi_{ba} \ra[ \xi_{bf}| \xi_{ba} \ra}
\f{[ \xi_{ac}| \xi_{ab}\ra[ \xi_{ad}| \xi_{ab} \ra}{[ \xi_{ac}| \xi_{ad}\ra}\,.
\label{eq:tau_chi}
\ee
Let $T_{ab}=\log \tau_{ab}$ with a chosen branch/lift. As an FN twist, $T_{ab}$ is conjugate to $2L_{ab}$ in the sense that $\{2L_{ab},T_{ab}\}=1$ and Poisson commutes with $\{2L_{cd}\}_{(cd)\neq(ab)}$ and $\{\cX_a,\cY_a\}$ (but not necessarily commutes with $\cT_{ab}$). 
This can be checked by using the framing flag definitions of FN length and FN twist. On the other hand, $T_{ab}$ can be obtained from the octahedron FG coordinates by symplectic transformation, which means $T_{ab}$ is a linear combination of the FG coordinates $(\vec{\Psi},\vec{\Pi})$ just as $(\vec{\fQ},\vec{\fP})$. We, therefore, conclude that $T_{ab}$ can be expressed in terms of the canonical pair $(2L_{ab},\cT_{ab})$ by linear transformation $T_{ab}=r\cdot 2L_{ab}+ \cT_{ab}+i\pi s$ with $r,s\in\R$. Such a relation makes sense also geometrically: any path on the annulus $(ab)$ can be approximated by a piecewise smooth path composed of meridian pieces, contributing some portion of $2L_{ab}$, and longitudinal pieces, contributing some portion of $T_{ab}$. Therefore, $\cT_{ab}$ corresponding the such a path can be expressed as a linear combination of $2L_{ab}$ and $T_{ab}$. $s$ comes from affine translation which does not affect the Poisson structure. 

For each given boundary condition of the 4-simplex, one can find two solutions $\fA$ and $\tfA$ to flat connections which correspond to opposite 4-volume of the 4-simplex, and they are related by parity transformation, analogous to the situation in the EPRL model \cite{Barrett:2009mw}. That is, $\sgn(V_4)|_{\fA}=-\sgn(V_4)|_{\tfA}$ \cite{Haggard:2014xoa,Han:2021kll}. Since $T_{ab}$ has dependence on $\sgn(V_4)$, these two flat connections in turn gives two solutions to $\cT_{ab}$: 
\be
\cT_{ab}|_{\fA}= -\f12\nu\sgn(V_4)\Theta_{ab}+i\pi N^{\fA}_{ab}+\zeta_{ab} \,,\quad
\cT_{ab}|_{\tfA}= \f12\nu\sgn(V_4)\Theta_{ab}+i\pi N^{\tfA}_{ab}+\zeta_{ab}\,,\quad
\zeta_{ab}=i\theta_{ab}+\f12\log\chi_{ab} -r \cdot 2L_{ab} +i\pi s\,,
\label{eq:cT_fA_tfA}
\ee
where $N^{\fA}_{ab},N^{\tfA}_{ab}\in\Z$ correspond to different lifts whose parities match as they correspond to the same $e^{\zeta_{ab}}$. It leads to the difference of the two momenta
\be
\cT_{ab}|_{\fA}-\cT_{ab}|_{\tfA}=-\nu\sgn(V_4)\Theta_{ab}+2\pi i N_{ab}\,,\quad
\text{with }
2N_{ab}=N^{\fA}_{ab}-N^{\tfA}_{ab}\in 2\Z\,.
\label{eq:cT_diff}
\ee

In summary, from the above derivation, we have seen that each FN length $2L_{ab}$ encodes the area of the triangle dual to the holes linked by the annulus $(ab)$ and that its dual FN twist $\cT_{ab}$ encodes the hyper-dihedral angle hinged by this triangle. Such a geometrical interpretation is useful  
for interpreting the critical solution to the equations of motion for the total amplitude \wrt the internal FN lengths in Section \ref{sec:critical_deficit}.

It remains to figure out the geometrical interpretation of $\theta_{ab}$ defined in \eqref{eq:def_tau}. 
Consider again the 4-vector representation $\bLambda_{ab}$ of $G_{ab}$ and its action on the triangle $f_{ab}$ shared by $T_a$ and $T_b$. the plane of $f_{ab}$ is spanned by the bivector $\star (N_a\w N_b)$ where $\star$ is the Hodge star operator. 
$\bLambda_{ab}$ changes the frame from $T_b$ to $T_a$. 
Consider a 4-vector ${V}$ that represents an edge of $f_{ab}$ shared by $T_a$ and $T_b$. $V$ is indeed in the plane of $\star (N_a\w N_b)$. 
$\bLambda_{ab}$ acts on the $V$ as 
(we omit the signs $\nu\sgn(V_4)$ in the following for conciseness) 
\be\begin{split}
\bLambda_{ab} V
&\equiv \bR_a  e^{-2\theta_{ab}\bJ^3}e^{-\Theta_{ab}\bK^3}\bR^{-1}_b  V
= \lb \bR_a e^{-2\theta_{ab}\bJ^3}\bR_b^{-1}\rb
\lb \bR_be^{-\Theta_{ab}\bK^3}\bR^{-1}_b\rb V 
=\lb \bR_a e^{-2\theta_{ab}\bJ^3}\bR_b^{-1}\rb  e^{-\Theta_{ab}\hat{n}_{ba}\cdot\vec{\bK} } V\,.
\end{split}
\label{eq:Lambda_act_bivector}
\ee
The boost generated by $e^{-\Theta_{ab}\hat{n}_{ba}\cdot\vec{\bK}}$ is along the normal $N_a\w N_b$ to $f_{ab}$ hence it keeps the plane spanned by the bivector $\star\lb N_a\w N_b\rb$, hence $V$ on the plane, invariant. 
Therefore, \eqref{eq:Lambda_act_bivector} can be simplified to be
\be
\bLambda_{ab} V
=\bR_a e^{-2\theta_{ab}\bJ^3}\bR_b^{-1}  V\,.
\ee 
$\bR_b^{-1}$ rotates the vector $\hat{z}$ to $-\hat{n}_{ba}$ in the frame of $T_b$, $e^{-2\theta_{ab}\bJ^3}$ generates a rotation around the $z$-axis, and $\bR_a$ rotates the vector $\hat{z}$ to $\hat{n}_{ab}$ in the frame of $T_a$. Therefore, in general, $V$ is rotated to a different vector by $\bLambda_{ab}$.

We are interested in a special case when the parallel transport is along a series of connected tetrahedra within the triangulation of a 4-manifold whose trajectory forms a (non-self-interacting) loop. That is, the initial and final tetrahedron in the transportation are the same: $T_a=T_b$, and we denote $G_{ab}=G_f$ and $\bLambda_{ab}=\bLambda_f$. In this case, $\theta_f$ can be determined in the following way.

Firstly, the rotation matrices $\bR_a=\bR_b=\bR$ in \eqref{eq:Lambda_act_bivector} as $\xi_{ab}=\xi_{ba}=\xi$. $\bLambda_f$ must keep the edge $V$ invariant since $f_{ab}$ remains the same, hence $\bR e^{-2\theta_{f}\bJ^3}\bR^{-1}  V\equiv e^{-2\theta_{f}\hat{n}\cdot \vec{\bJ}}\stackrel{!}{=} V$. where $\hat{n}=\hat{n}_{ab}=\hat{n}_{ba}$.  
  Since $e^{-2\theta_{f}\hat{n}\cdot \vec{\bJ}}$ generates a rotation around the normal $\hat{n}$ to $f_{ab}$ by an angle $-2\theta_{f}$, 
  $V$ is kept invariant only when (recall the range $\theta_{f}\in[0,2\pi)$)\footnote{This result was also derived in Appendix F of \cite{Han:2015gma} in a slightly different manner.}
\be
2\theta_{f}=0 \,\text{ or }\, 2\pi
\quad\Longleftrightarrow\quad
\theta_{f}=0\,\text{ or }\, \pi\,.
\label{eq:zero_deficit}
\ee
Returning to the fundamental representation \eqref{eq:def_tau}, the choice $\theta_{f}=\pi$ changes $\gamma_{f}=\gamma_{ab}$ to $\gamma_f^{-1}$ compared to the choice $\theta_{f}=0$. The two solutions to $\theta_{f}$ can be understood as different lifts from $\SO(1,3)^+$ to $\SL(2,\bC)$. In other words, if we interpret the lift $\theta_{f}=0$ as a time-oriented map $\SO(1,3)^+\rightarrow \SO(1,3)^+$, then the lift $\theta_{f}=\pi$ can be interpreted as a time-flipping map $\SO(1,3)^+\rightarrow \SO(1,3)^-$. Such an interpretation makes sense because the  
time-like normal to a tetrahedron on the boundary of a 4-simplex can be future-pointing or past-pointing. When the 4-manifold, hence its triangulation, is globally time-oriented, the unique solution $\theta_{f}=0$ for all $f$'s is picked. 
We see in Section \ref{sec:critical_deficit} that such a solution leads to the uniqueness of the solution to the deficit angle.

\section{Fock-Goncharov coordinates on $\cS_2$ and $\cS'_1$ in $\Delta_3$ 4-complex}
\label{app:glue}

Denote the framing flag parallel transported from hole $i$ of $\cS_2$ (\resp $\cS'_1$) to a common point on $\cS_2$ (\resp $\cS'_1$) as $s_i$ (\resp $s'_i$). 
Denote the edge of the ideal triangulation of $\cS_2$ (\resp $\cS'_1$) connecting hole $i$ and hole $j$ as $e_{ij}$ (\resp $e'_{ij}$). Then the FG coordinates on the 6 edges are summarized as follows.
\be
\ba{clccl}
 e_{13}: & e^{\cX^{(2)}_{14}}=e^{-\cY_2}=\f{\la s_4 \w s_1 \ra\la s_3 \w s_2 \ra}{\la s_4 \w s_3 \ra\la s_1 \w s_2 \ra} &\quad &
 e'_{12}: & e^{\cX^{'(1)}_{25}}=e^{\cX'_1}=\f{\la s'_3 \w s'_1 \ra\la s'_2 \w s'_4 \ra}{\la s'_3 \w s'_2 \ra\la s'_1 \w s'_4 \ra}=\f{\la s_2 \w s_1 \ra\la s_3 \w s_4 \ra}{\la s_2 \w s_3 \ra\la s_1 \w s_4 \ra}\equiv e^{\cY_2}\\[0.15cm]
 e_{12}: & e^{\cX^{(2)}_{15}}=e^{\cX_2}=\f{\la s_3 \w s_1 \ra\la s_2 \w s_4 \ra}{\la s_3 \w s_2 \ra\la s_1 \w s_4 \ra} &\quad &
 e'_{13}: & e^{\cX^{'(1)}_{23}}=e^{-\cY'_1}=\f{\la s'_4 \w s'_1 \ra\la s'_3 \w s'_2 \ra}{\la s'_4 \w s'_3 \ra\la s'_1 \w s'_2 \ra}=\f{\la s_4 \w s_1 \ra\la s_2 \w s_3 \ra}{\la s_4 \w s_2 \ra\la s_1 \w s_3 \ra}\equiv e^{-\cX_2}\\[0.15cm]
  e_{14}: & e^{\cX^{(2)}_{45}}=\f{\la s_2 \w s_1 \ra\la s_4 \w s_3 \ra}{\la s_2 \w s_4 \ra\la s_1 \w s_3 \ra} &\quad &
 e'_{14}: & e^{\cX^{'(1)}_{35}}=\f{\la s'_2 \w s'_1 \ra\la s'_4 \w s'_3 \ra}{\la s'_2 \w s'_4 \ra\la s'_1 \w s'_3 \ra}=\f{\la s_3 \w s_1 \ra\la s_4 \w s_2 \ra}{\la s_3 \w s_4 \ra\la s_1 \w s_2 \ra}\equiv e^{-\chi^{(2)}_{45}}\\[0.15cm]
   e_{23}: & e^{\cX^{(2)}_{13}}=\f{\la s_1 \w s_2 \ra\la s_3 \w s_4 \ra}{\la s_1 \w s_3 \ra\la s_2 \w s_4 \ra} &\quad &
 e'_{23}: & e^{\cX^{'(1)}_{24}}=\f{\la s'_1 \w s'_2 \ra\la s'_3 \w s'_4 \ra}{\la s'_1 \w s'_3 \ra\la s'_2 \w s'_4 \ra}=\f{\la s_1 \w s_3 \ra\la s_2 \w s_4 \ra}{\la s_1 \w s_2 \ra\la s_3 \w s_4 \ra}\equiv e^{-\chi^{(2)}_{13}}\\[0.15cm]
   e_{24}: & e^{\cX^{(2)}_{35}}=\f{\la s_3 \w s_2 \ra\la s_4 \w s_1 \ra}{\la s_3 \w s_4 \ra\la s_2 \w s_1 \ra} &\quad &
 e'_{34}: & e^{\cX^{'(1)}_{34}}=\f{\la s'_1 \w s'_3 \ra\la s'_4 \w s'_2 \ra}{\la s'_1 \w s'_4 \ra\la s'_3 \w s'_2 \ra}=\f{\la s_1 \w s_2 \ra\la s_4 \w s_3 \ra}{\la s_1 \w s_4 \ra\la s_2 \w s_3 \ra}\equiv e^{-\chi^{(2)}_{35}}\\[0.15cm]
   e_{34}: & e^{\cX^{(2)}_{34}}=\f{\la s_1 \w s_3 \ra\la s_4 \w s_2 \ra}{\la s_1 \w s_4 \ra\la s_3 \w s_2 \ra} &\quad &
 e'_{24}: & e^{\cX^{'(1)}_{45}}=\f{\la s'_3 \w s'_2 \ra\la s'_4 \w s'_1 \ra}{\la s'_3 \w s'_4 \ra\la s'_2 \w s'_1 \ra}=\f{\la s_2 \w s_3 \ra\la s_4 \w s_1 \ra}{\la s_2 \w s_4 \ra\la s_3 \w s_1 \ra}\equiv e^{-\chi^{(2)}_{34}}
\ea\,.
\label{eq:glue_FG}
\ee
{One finds that, from the calculation point of view, the gluing of 4-holed spheres follows the same way as gluing ideal tetrahedra to form an ideal octahedron ({it r.f.} Section \ref{sec:review}). This is because, although we need to flipped the orientation of $\cS'_1$, we read the labels of holes on the quadrilateral (lower panel of fig.\ref{fig:glue}) from the ``inside'' of $\cS'_1$. Then this is the same as reading the labels from the ``outside'' before flipping the orientation of $\cS'_1$.}

\section{Proof of Theorem \ref{theorem:hormander}}
\label{app:proof}

$\sum_{f=1}^F \lb \f{i}{2\pi} \cF_f(2L_f)-2u_fL_f\rb$ in \eqref{eq:effective_action} comes from the face amplitudes and it is only imaginary since $\cF(2L_f)$ is a real function of $2L_f$ upon the imposition of simplicity constraints. 
 We are left to consider each $S^v_{\vec{p}^v,\vec{u}^v,\vec{\hrho}^v}$ \eqref{eq:effective_action_all} obtained from the large-$k$ approximation of partition function \eqref{eq:partition_S3G5} and coherent states \eqref{eq:coherent_state_def} for a spinfoam vertex. 
We first observe that the positive angles that contribute to the imaginary parts of $\{\mu_I,\nu_I\}$ are not seen at the large-$k$ approximation of the action. Then each tilted variable is merely the complex conjugate of its non-tilted counterpart. Additionally, $b^{-1}$ is the complex conjugate of $b$ as it is a phase. Therefore, $S_1^v+\widetilde{S}^v_1$ is pure imaginary obviously seen from their expressions \eqref{eq:S1} -- \eqref{eq:S1t}. We next consider the rest of the first line of \eqref{eq:effective_action_all}, which can be rewritten as 
\be
S_0^v
-\f{2\pi i}{k} \vec{p}^v\cdot\vec{n}^v
=\f{\pi i}{k^2}\left[-2\left(\vec{\mu}^v-\frac{i Q}{2} \vec{t}\right) \cdot \vec{\nu}^v+2 \vec{m}^v \cdot \vec{n}^v-\vec{\nu}^v \cdot \bA\bB^\top \cdot \vec{\nu}^v+ \vec{n}^v \cdot \bA \bB^\top \cdot \vec{n}^v+k\vec{n}^v\cdot( \vec{t}+
2\vec{p}^v)\right]\,.
\label{eq:S0_old_coord}
\ee
$\vec{\mu}^v,\,\vec{\nu}^v$ can be viewed as real variables at large $k$ hence the above expression is also pure imaginary. The second line of \eqref{eq:effective_action_all} contains the logarithms of coherent states and a term $\f{2\pi i}{k}\sum_{a=1}^5u_a^vm_a^v$ from the Poisson resummation for $m_a^v$. The latter is apparently imaginary. All the real parts of \eqref{eq:effective_action}, therefore, come from the coherent states. 
Due to the nature of coherent states (and is clear from the definitions \eqref{eq:coherent_state_def}), the norms are Gaussian and hence must contribute a non-positive real part for the action with zero obtained at the critical point. This proves the first two equations of \eqref{eq:action_property}. The last equation is the definition of a critical point hence is trivially satisfied. 

The first two equations of \eqref{eq:action_property} also imply that the real parts of the eigenvalues of the Hessian, denoted as $\re(H_{\vec{x}})$, satisfy $\re(H_{\vec{x}})\leq 0$ at the neighbourhood of the real critical point $\vec{x}_0$. 

As $S(\vec{x},\vec{r})$ is apparently analytic near $\vec{x}_0$, its analytic continuation $S(\vec{z},\vec{r})$ is also analytic near the complex critical point $\vec{z}_0(\vec{r})$. Then $S(\vec{z},\vec{r})$ possess a convergent Taylor series at $\vec{z}_0(\vec{r})$:
\be
S(\vec{z},\vec{r}) = S(\vec{z}_0(\vec{r}),\vec{r}) +\sum_{|\alpha|=2}\f{1}{\alpha!}\left.D^\alpha S(\vec{z},\vec{r})\right|_{\vec{z}=\vec{z}_0(\vec{r})}(\vec{z}-\vec{z}_0(\vec{r}))^\alpha + O(|\vec{z}-\vec{z}_0(\vec{r})|^3)\,,
\label{eq:Taylor}
\ee
where $D^\alpha$ stands for the derivative of order $\alpha$ acting on a function $f(\vec{z})$ with $\vec{z}\in\bC^n$ as
\be
D^\alpha f =\f{\partial^{|\alpha|}f }{\partial z_1^{\alpha_1}\cdots \partial z_n^{\alpha_n} }\,,\quad
|\alpha|=\alpha_1+\cdots+\alpha_n
\ee
and $\alpha!:=\alpha_1!\cdots\alpha_n!$.
$D^\alpha S(\vec{z},\vec{r})$ with $|\alpha|=2$ is simply the Hessian $H_{\vec{z}}$ of the action. 

As assumed, the complex critical point $\vec{z}_0$ is in the neighbourhood $U$ of the real critical point $\vec{x}_0$, as illustrated in fig.\ref{fig:complex_critical_point}. 
$\vec{z}_0(\vec{r})$ is an analytic function in $\vec{r}$. Let $\vec{x}_0(\vec{r})=\vec{z}_0(\vec{r}_0)$. Then $\vec{z}_0(\vec{r})$ can be viewed as a path in $\bC^n$ starting at $\vec{x}_0$. 
Within the neighbourhood $U$, $\re(H_{\vec{x}})\leq 0$ implies $\re(H_{\vec{z}_0})\leq 0$ by analyticity, which in turn implies $S(\vec{z}_0, \vec{r})\leq 0$. By \eqref{eq:Taylor}, we have
\be
\re\lb S(\vec{z}_0, \vec{r})\rb + \sum_{|\alpha|=2}\f{1}{\alpha !}\re\lb H_{\vec{z}_0}\lb\vec{z}-\vec{z}_0\rb^\alpha\rb+ \re\lb O(|\vec{z}-\vec{z}_0|^3\rb\leq0\,.
\label{eq:Re_Taylor}
\ee
Consider $\vec{z}=\re(\vec{z}_0)+|\im(\vec{z}_0)|\vec{s}$ with some $\vec{s}\in\R^n\,,\,|\vec{s}|<1$. When $\im(\vec{z}_0)$ is small, $\vec{z}$ parametrized in this way is within $U$ hence \eqref{eq:Taylor} is valid. Define $\vec{\eta}=\im(\vec{z}_0)/|\im(\vec{z}_0)|$. Then \eqref{eq:Re_Taylor} can be reformulated as
\be
\re(S(\vec{z}_0,\vec{r}))\leq -|\im(\vec{z}_0)|^2\lb \sup_{|\vec{s}|<1} \sum_{|\alpha|=2}\f{1}{\alpha !} \re\lb H_{\vec{z}_0} (\vec{s}-i\vec{\eta})^\alpha \rb+ C'|\im(\vec{z}_0)|\rb\,,
\label{eq:Re_Taylor_2}
\ee
where $0<C'<\infty$ is some real constant. We are left to prove that the expression in the bracket above is non-negative (as it is indeed bounded). Firstly, $\re(H_{\vec{z}_0})\leq0$ as observed above. We expand the term $\sum_{|\alpha|=2}\f{1}{\alpha!}\re\lb H_{\vec{z}_0}(\vec{s}+i\vec{\eta})^\alpha\rb$:
\be
\f12\sum_{i,j=1}^n \left[ \re(H_{\vec{z}_0})_{ij}\lb s_is_j - \eta_i\eta_j\rb - 2\im(H_{\vec{z}_0})_{ij}s_i\eta_j\right]=:\f12\la \vec{s},\re(H_{\vec{z}_0})\vec{s}\ra - \f12\la \vec{\eta},\re(H_{\vec{z}_0})\vec{\eta}\ra-\la \vec{s},\im(H_{\vec{z}_0})\vec{\eta}\ra  \,.
\label{eq:2-form_expand}
\ee
To proceed, we only need to find an admissible $\vec{s}$ ($|\vec{s}|<1$) such that \eqref{eq:2-form_expand} is positive. To this end, if $\la \vec{\eta},\re(H_{\vec{z}_0})\vec{\eta}\ra\neq 0$, we let $\vec{s}=0$. Then \eqref{eq:Re_Taylor_2} is proven as $\la \vec{\eta},\re(H_{\vec{z}_0})\vec{\eta}\ra< 0$ is guaranteed by $\re(H_{\vec{z}_0})\leq 0$. If $\la \vec{\eta},\re(H_{\vec{z}_0})\vec{\eta}\ra=0$, then $\re(H_{\vec{z}_0})=\vec{0}$. The assumption $\det(H_{\vec{z}_0})\neq 0$ then implies that $\im(H_{\vec{z}_0})\neq \vec{0}$. In this case, we take $\vec{s}=-\epsilon\im(H_{\vec{z}_0})\vec{\eta}$ with $\epsilon>0$ being small so that $|\vec{s}|<1$ is not violated. Then $\la \vec{s},\im(H_{\vec{z}_0})\vec{\eta}\ra >0$ hence \eqref{eq:2-form_expand} is positive. Therefore, \eqref{eq:decay} is proved. 
\eqref{eq:stationary_phase} is the result from stationary analysis with distinct critical points added. We refer to Theorem 7.7.12 in \cite{hormander2015analysis} for a detailed proof. The distinctness of critical points is implied by $\det(H_{\vec{z}_0})\neq0$ as, otherwise, continuous critical points would lead to degenerate directions of the Hessian.

\bibliographystyle{bib-style} 
\bibliography{CDA.bib}

\end{document}